\documentclass[journal]{IEEEtran}

\ifCLASSINFOpdf
\else
\fi

\usepackage[cmex10]{amsmath}
\ifCLASSOPTIONcompsoc
\usepackage[caption=false,font=normalsize,labelfont=sf,textfont=sf]{subfig}
\else
\usepackage[caption=false,font=footnotesize]{subfig}
\fi

\usepackage{fancybox}
\usepackage{color}
\fancyput(0cm,2cm){\textcolor{black}{\doublebox{%
\href{https://doi.org/10.1109/TBCAS.2018.2881219}{DOI: 10.1109/TBCAS.2018.2881219}, \copyright~2018~IEEE. The article has been accepted for publication.%
}}}
\usepackage{amssymb}
\usepackage{amsmath}
\usepackage{array}
\usepackage{cite}
\usepackage{color}
\usepackage[]{graphicx}
\usepackage{nccmath}
\usepackage{nidanfloat}
\usepackage[percent]{overpic}
\usepackage{pifont}
\usepackage{subfig}
\usepackage{tabularx}
\usepackage{tikz}
\usepackage{upgreek}
\usepackage{url}
\usepackage{verbatim}

\input epsf
\newcommand{\fulltoday}{\ifcase\month\or
    January\or February\or March\or April\or May\or June\or
    July\or August\or September\or October\or November\or December\fi
    \space\number\day\space \number\year}
\def\cw{\columnwidth}
\def\tw{\textwidth}

\def\iic{I\textsuperscript{2}C}
\def\iiw{I\textsuperscript{2}We}
\def\hi{{\rm H}}
\def\lo{{\rm L}}
\def\in{{\rm in}}
\def\Zio{Z_{\rm IO}}
\def\Zpo{Z_{\rm P1}}
\def\Zpt{Z_{\rm P2}}
\def\Cioh{C_{\rm H}}
\def\Rioh{R_{\rm H}}
\def\Riol{R_{\rm L}}
\def\Vf{V_{\rm F}}
\def\Vh{V_1^{\rm H}}
\def\Vl{V_1^{\rm L}}
\def\Xioh{X_{\rm IO}^{\rm H}}
\def\Zinh{Z_{\rm in}^{\rm H}}
\def\Zinl{Z_{\rm in}^{\rm L}}
\def\Zp{Z_{\rm P}}
\def\zm{z_{\rm m}}
\def\xm{x_{\rm m}}

\def\Lma{L_{\rm ma}}
\def\Lmb{L_{\rm mb}}
\def\La{L_{1{\rm a}}}

\def\Lb{L_{1{\rm b}}}
\def\LLb{L_{2{\rm b}}}
\def\Ca{C_{1{\rm a}}}
\def\CCa{C_{2{\rm a}}}
\def\Cb{C_{1{\rm b}}}

\def\stop{{\rm stop}}
\def\mod{{\rm mod}}
\def\j{{\rm j}}

\def\im#1{\operatorname{Im}({#1})}

\newcommand{\reffig}[1]{Fig. \ref{fig:#1}}
\newcommand{\reftab}[1]{Table \ref{tab:#1}}
\newcommand{\refeq}[1]{(\ref{eq:#1})}
\newcommand{\refsec}[1]{Section \ref{sec:#1}}
\newcommand{\refssec}[1]{Section \ref{ssec:#1}}

\begin{document}
\title{Inter-IC for Wearables (\iiw ): Power and Data Transfer over Double-sided Conductive Textile}

\author{Akihito~Noda,~\IEEEmembership{Member,~IEEE}
and~Hiroyuki~Shinoda,~\IEEEmembership{Member,~IEEE}
\thanks{Manuscript received August 5, 2018; revised November 1, 2018. %
This work was supported in part by JSPS KAKENHI Grant Number 17H04685. 
Double-sided conductive textiles were provided by Teijin Limited, Tokyo, Japan.
}%
\thanks{Supplementary downloadable material related to this paper is available at http:/{\slash}ieeexplore.ieee.org, provided by the authors. 
This is a multimedia MP4 format movie clip, which shows a demonstration of an \iiw\ network with temperature sensors. 
This material is 18.9 MB in size.}
\thanks{A. Noda is with the Department of Mechatronics, Nanzan University, Nagoya 466-8673, Japan (e-mail: anoda@nanzan-u.ac.jp).}
\thanks{H. Shinoda is with the Department of Complexity Science and Engineering, The University of Tokyo, Chiba 277-8561, Japan (e-mail: hiroyuki\_shinoda@k.u-tokyo.ac.jp).}
}
\markboth{IEEE TRANSACTIONS ON BIOMEDICAL CIRCUITS and SYSTEMS}%
{Noda \MakeLowercase{\textit{et al.}}: \iiw: POWER AND DATA TRANSFER OVER DOUBLE-SIDED CONDUCTIVE TEXTILE}

\maketitle

\begin{abstract}
We propose a power and data transfer network on a conductive fabric material based on an existing serial communication protocol, Inter-Integrated Circuit (\iic). 
We call the proposed network inter-IC for wearables (\iiw).
Continuous dc power and \iic-formatted data are simultaneously transferred to tiny sensor nodes distributed on a double-sided conductive textile.
The textile comprises two conductive sides, isolated from each other, and is used as a single planar transmission line.
\iic\ data are transferred along with dc power supply based on frequency division multiplexing.
Two carriers are modulated with the clock and the data signals of \iic.
A modulation and demodulation circuit is designed such that off-the-shelf \iic-interfaced sensor ICs can be used. 
The novelty of this work is that a special filter to enable passive modulation is designed by locating its impedance poles and zeros at the appropriate frequencies. 
{The term ``passive modulation'' herein implies that the sensor nodes do not generate carrier waves by themselves; instead, they reflect only the externally supplied careers for modulation.}
The proposed scheme enables the flexible implementation of wearable sensor systems in which multiple off-the-shelf tiny sensors are distributed throughout the system.
\end{abstract}


\IEEEpeerreviewmaketitle

\section{Introduction}

\IEEEPARstart{S}{mart}
fabrics, or e-textiles \cite{etextileJPROC2003}, 
that can perform biomedical sensing on a user's body surface while eliminating a number of individual wires is a key technology for wearable health monitor systems.
Distributing various sensors throughout for temperature measurement, accelerometers, electrocardiography/electromyography, etc., will yield rich information for health care \cite{khanADMA2016} and is applicable to human--computer interfaces \cite{georgi2015recognizing}.

Powering the distributed sensors and gathering the sensor readings on {the system are critical for} implementing such wearable sensor systems. 
Although functionalizing a fiber itself, e.g., piezoelectric fibers \cite{egusa2010multimaterial,tajitsuTDEI2015} and strain-sensing fibers \cite{ZHANG2006129,chunyaADMA2016}, is a promising approach to embed sensors into clothing without sacrificing comfort, an entire system still requires some silicon-based circuit components to include amplifiers and digital communication interfaces. 
Therefore, the integration of textiles and silicon-based circuitry, which are produced separately, is required to build a complete wearable sensing system. 
The motivation of this work is to present a practical method to integrate off-the-shelf silicon-based sensors and textiles. 

One modern, straightforward approach is radio-based wireless communication. 
Each sensor operates with a tiny battery and sends data over the air. 
However, {frequent data transmission by densely distributed sensors will involve significant difficulties including interference, latency, and battery life.} 

Another approach to avoid these difficulties is wire-based communication. 
For the physical connection in wearable systems, circuit patterning on fabric with 
conductive yarns %
\cite{buechleyPervasive2008
,Li20140472}, 
conductive screen printing ink %
\cite{kimTADVP2010
,matsuhisa2015printable}, 
or ironed-on conductive fabric patches \cite{Buechley2009}, 
instead of copper tracks on conventional printed circuit boards (PCBs), is a straightforward and practical approach. 
This approach requires separate conductive lines connecting the leads/pads of electronic components individually, as in conventional PCBs. 
A possible vulnerability is that the fraying/whiskering of the fabric would cause a fatal breakdown, such as the disconnection of tracks and short-circuit between neighboring conductive tracks.

An alternative approach to such one-to-one wiring is by connecting all the devices on the clothing with a single bus. 
A double-sided conductive fabric sheet, i.e., both sides of the sheet are conductive and are insulated from each other, can be used as the bus \cite{learhoven2003,wadeMPRV2007,akita2006flexible,shinoINSS2007,tajimaAH2016,nodaIMS2017}.

The most significant advantage of this approach is that the conductive textile is patterned uniformly without depending on the sensor node locations. 
Therefore, the textile and electronic circuitry can be designed and fabricated independently from each other.
The same textile is typically used in various products and is cost effective.
Because the sensor nodes do not require being located precisely  on the textile, they can be detached and reassembled in daily use;  subsequently, the wear is washable even if the sensors are not waterproof. 
It is noteworthy that the uniform pattern is not mandatory, and any other artwork patterns with conductive yarns/ink can be used if they are preferred. 

As a simple implementation using commercially available iButton \cite{ibutton} devices, the MicroLAN \cite{DS9091K} sensor network on a layered conductive fabric has been presented in \cite{learhoven2003}.
TextileNet \cite{akita2006flexible} enables a dc power supply and serial communication on the transmission line by temporally switching the power supply state and the pulse transfer state; thus, the power supply is intermittent and only a singleplexed signal transfer is supported. 
Wade and Asada proposed a dc powerline communication on a layered conductive textile \cite{wadeICRA2004,wadeMPRV2007}, using a modulation/demodulation (modem) circuitry developed by themselves \cite{wadeDCPLC2006}. 
A continuous dc power supply and amplitude modulated signal using a carrier in a frequency range of 2--10 MHz are transferred simultaneously on the same textile medium. 
Based on a similar concept, continuous dc power supply combined with multiplexed signal transfer \cite{nodaIMS2017} derived from multiplexed ac power transfer \cite{tajimaAH2016}, have been presented, based on frequency division multiplexing (FDM). 
In those works, the three-channel on/off control of slave devices was demonstrated by the on-off modulation of three carriers. 
This scheme requires $N$ carriers at different frequencies to achieve $N$-channel on-off control, i.e., $N$-bit signal transfer. 

Herein, the FDM scheme is utilized to transfer a pair of clock and data signals of a serial communication protocol, i.e., Inter-Integrated Circuit (\iic) \cite{semiconductorsum10204}.
It provides compatibility with a significantly wider range of commercially available sensor ICs and microcontrollers (MCUs), compared with iButton devices or specially designed modems presented in the previous works mentioned above. 

\begin{figure}[t]
	\centering
	\includegraphics[width=0.9\cw]{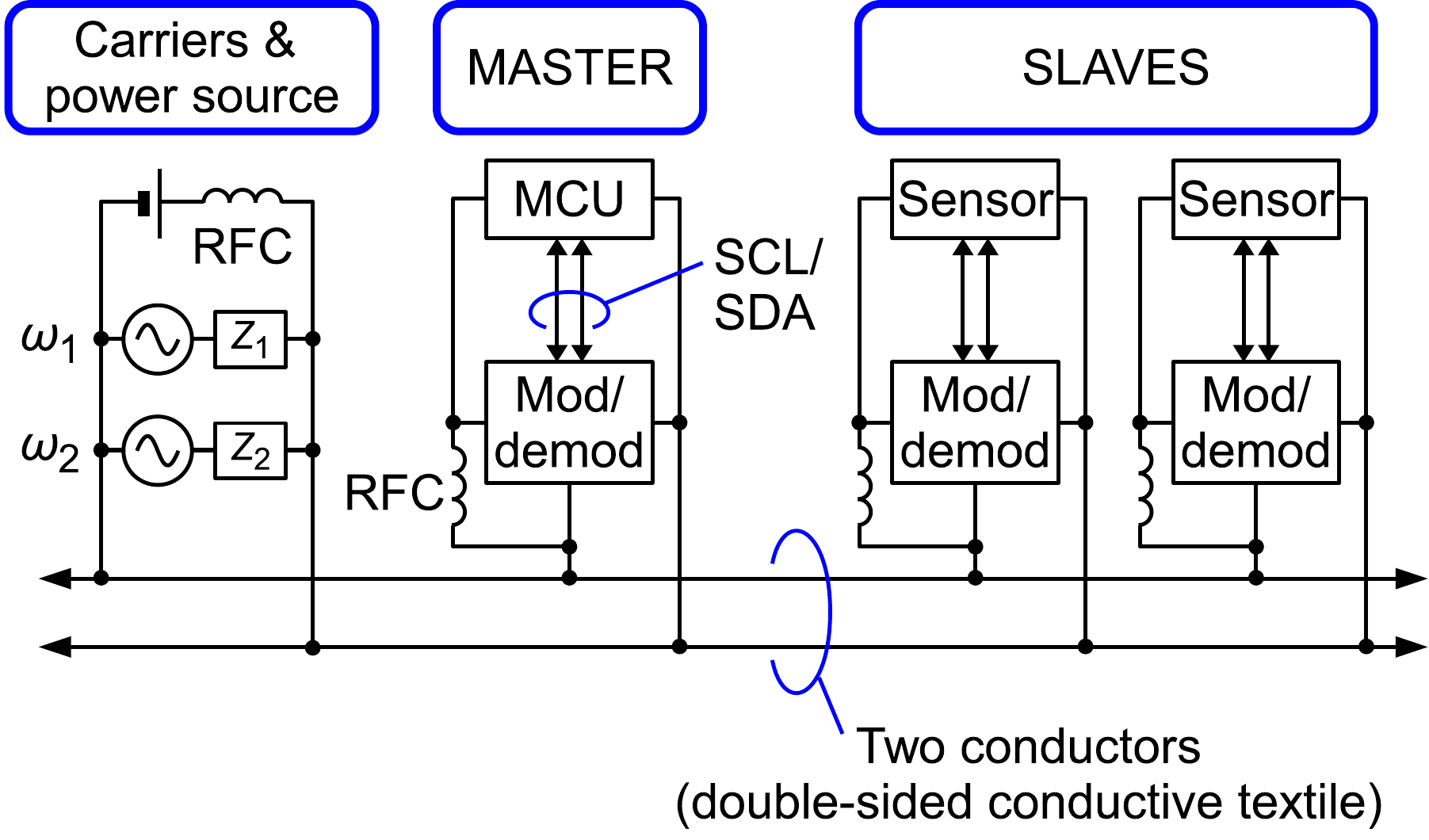}
	\caption{Schematic diagram of proposed \iiw\ system.}
	\label{fig:fdm}
\end{figure}

The primary contribution of this study is the proposal of inter-IC for wearables (\iiw) that combines \iic\ data transfer and dc power supply on a double-sided conductive textile. 
As illustrated in \reffig{fdm}, the dc power and \iic\ signals are transferred simultaneously on the same planar transmission line using FDM, similar to other PLCs \cite{wadeDCPLC2006,ferreiraAFRICON1996,liuICRA2001}.
An important advantage of the proposed method is the small footprint and low power consumption achieved by passive modulation.
\iic-interfaced ICs send the output data by modulating the carriers supplied externally, similar to passive radio frequency identification (RFID) \cite{raoMAP2006} and other backscatter communications \cite{bharadiaSIGCOMM2015,kelloggNSDI2016,zhangSIGCOMM2016}.
{Indeed, the modulation operation relies on an ``active'' digital IC; the term ``passive modulation'' herein implies, nonetheless, that the}
carrier generator is separated from the modulation circuitry and eliminated from each tiny sensor node.
A specially designed LC filter, described in \refsec{principle}, enables the passive modulation. 
The filter design considering the location of its impedance poles and zeros at appropriate frequencies is the novelty of this work. 
{The theoretical analysis on the filter design and comparison with simulation/experiment are additional contributions of this paper to our previous conference paper \cite{nodaBSN2018}.}
The carriers are modulated directly with the \iic\ interface outputs of a digital IC, and also demodulated with a simple carrier detector and a data slicer without any additional digital signal processor or buffer memories.

Contrary to conventional radio links, the proposed \iiw\ link can eliminate the latency due to listen-before-talk, and collision followed by retransmission. 
In radio-based wireless sensor networks, such latency causes the out-of-order arrival of data from distributed sensor nodes, and disturbs processing such as sensor fusion at the sink node.
To recover the order of the data as they are acquired by sensors, a time base must be shared in the sensor network for time stamping, in which additional resources are required on each node.
In an \iiw\ network, as well as \iic, the master fully initiates data transfer transactions and receives sensor data in order; therefore, no additional processing for data ordering is required.

Another important feature from a practical viewpoint is, as stated above, that the \iiw\ network is implemented with off-the-shelf \iic-enabled ICs and MCUs.
Engineers familiar with the \iic\ network can easily reuse the same ICs and MCUs, as well as software libraries implemented in MCUs. 
Additionally, some of the commercially available low-cost sensor ICs, e.g., inertial measurement units (IMUs), provide well-engineered sensor fusion functionalities integrated into a single chip. 
\iiw\ enables reusing such functions and provides an option to avoid reinventing the same functionalities. 

{\iiw\ enables sensor-distributed clothing. 
Potential applications include distributed body temperature sensors for medical monitoring/diagnosis \cite{webb2013ultrathin}, and distributed posture sensors for motion analyses in rehabilitation \cite{sardiniTIM2015}.
Multimodal sensing systems can also be implemented, because various types of \iic-interfaced sensors can coexist on the same clothing.}

The remainder of the paper is organized as follows.
\refsec{textile} briefly describes the structure of a double-sided conductive textile and a tack connector. 
\refsec{principle} explains the operation principle of the proposed \iiw\ network.
A design example and a demonstration system are presented in \refsec{example} and \refsec{demo}, respectively.
Finally, conclusions are presented in \refsec{conclusion}.

\section{Double-sided Conductive Textile}
\label{sec:textile}

This section briefly describes the double-sided conductive textile and a tack connector used in this work. 
The proposed \iiw\ communication is applicable to any two-conductor transmission lines including coaxial, twisted-pair, and parallel-pair cables. 
Nonetheless, combination with a double-sided conductive textile would be suitable for wearable systems, because it provides the freedom of sensor location along with a lower resistance and higher current capacity, while using a fairly flexible conductive textile for comfortable wearability.
The advantages of such two-dimensional (2-D) conductive material over a one-dimensional (1-D) conductive strand are discussed in detail in \cite{wadeMPRV2007}.
The resistance between two points on the 2-D material is approximately proportional to the logarithm of the distance between the two points, while the resistance of a 1-D conductor is proportional to its length. 
{In this work, we assume that the resistance is almost frequency independent, because the silver plating thickness of the conductive yarns is significantly smaller than the skin depth at the carrier frequencies used in this work.
The resistance is also stable with respect to the deformation of the textile.}
Additionally, on the 2-D material, the current paths are redundant, and the connection is retained robustly even if the material is partially damaged. 

\begin{figure}[t]
	\centering
	\includegraphics[width=0.7\cw]{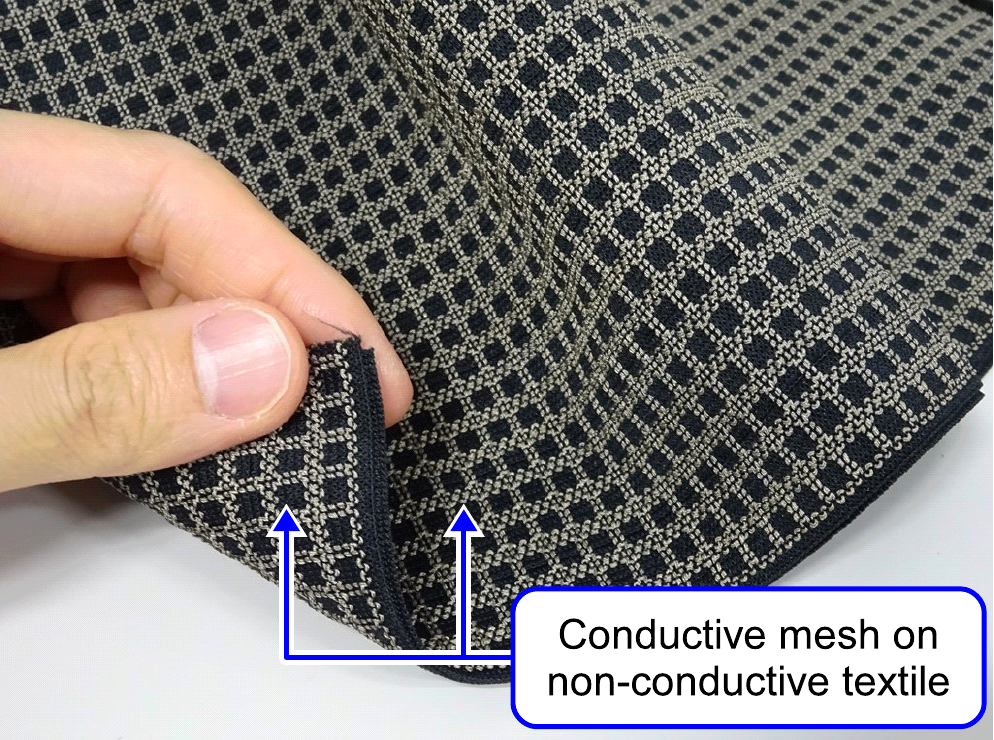}
	\caption{Example of double-sided conductive textile. 
		Square mesh patterns are formed on both the sides of the 
		base fabric with silver-plated conductive yarns. }
	\label{fig:sheetphoto}
\end{figure}

\begin{figure}[t]
	\centering
	\includegraphics[width=0.8\cw]{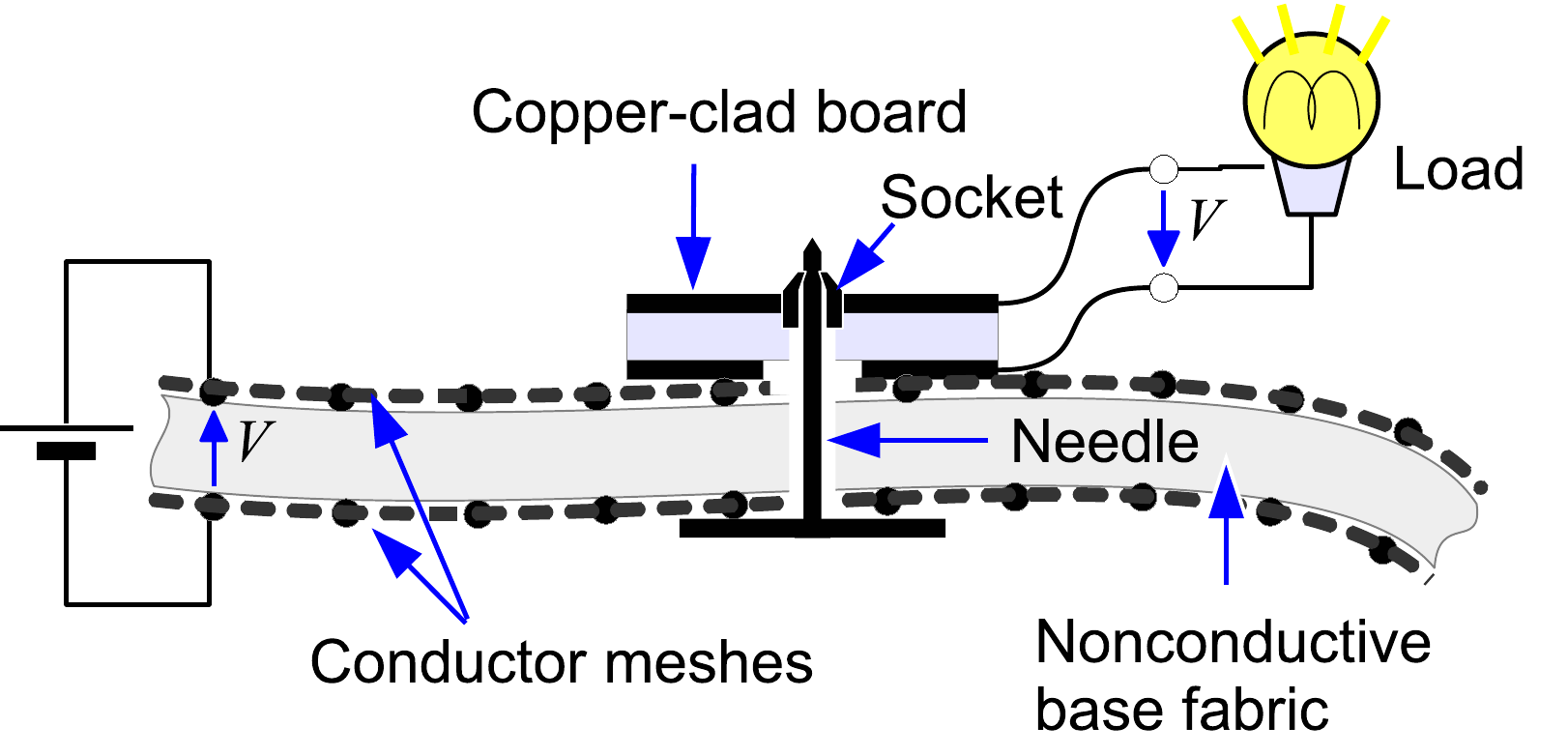}
	\caption{Cross-sectional view of a wearable signal/power transfer sheet and a tack connector.
		Although the mesh conductors are represented with separate dots 
		(and a dotted line connecting them) in this figure,  
		each of them is electrically continuous. 
		The top and bottom mesh conductors are 
		isolated from each other. }
	\label{fig:crosssection}
\end{figure}

An example of the textile is shown in \reffig{sheetphoto}.
The base textile is nonconductive; further, on both of its sides, square mesh patterns are embroidered with conductive yarns. 
A cross-sectional view of the structure is illustrated in \reffig{crosssection}.
The conductive yarns on one side are electrically connected together and isolated from the conductive mesh on the other side.
The square mesh is not mandatory and any other artwork patterns can be used. 
Alternately, conductive yarns can be integrated at the stage of the textile fabrication, i.e., the textile can be woven/knitted with both conductive and nonconductive yarns.
Piling up conductive and nonconductive textile sheets or printing conductive ink on both the sides of a nonconductive base textiles sheet are also acceptable.
Regardless of the fabrication method, an essential requirement is that the conductive materials are exposed on both sides, and both the conductive surfaces are isolated from each other.

As shown in \reffig{crosssection}, a special connector consisting of a tack (needle) and a socket is used in this work. 
The back side of the PCB is in direct contact with one side of the textile, and the conductive needle-and-socket connector enables the electrical connection between the PCB and the other side of the textile.
To avoid shorting both conductive surfaces, the needle should be partially insulated or placed in a nonconductive area, i.e., inside an aperture of the mesh. 
This configuration allows for the physical mounting of a device, and its electrical connection to be integrated into a single action, i.e.,  sticking the needle through the textile and mating the needle with the socket. 
{%
The pinching of the textile induces only a negligible effect on the impedance of the textile.}

{%
The voltage applied between the two conductive planes will also be applied to the user's body, if the user touches both naked conductive surfaces simultaneously. 
Provided that the voltage is as low as the standard power rail voltage for modern digital circuitry, such as 3.3 V, the voltage will not be harmful for the user. 
Nonetheless, if the system is required to avoid applying the voltage on the user's body, the conductive surfaces must be covered with the appropriate insulation layers. }

\section{Principle of Operation}
\label{sec:principle}

\subsection{Inter-Integrated Circuit (\iic)}

This subsection briefly explains the operation and implementation of the \iic. 
The descriptions below assume the standard mode or fast mode, with clock frequencies up to 100 kHz or 400 kHz, respectively, defined in the \iic\ specification \cite{semiconductorsum10204}. 

An \iic\ bus consists of two signal lines: clock (SCL) and data (SDA). 
One or more masters, and one or more slaves can be connected to the bus in parallel.
Only one master can initiate the communication at a time, and all the other slaves on the bus listen to the master's signal; subsequently, only a slave addressed by the master responds to the signal during the transaction. 
The bus is bidirectional, i.e., the data can be transferred from the master to the slave, and vice versa.
For both directions, the master generates the clock signal, and thus initiates the communication. 
The data signal is generated by the master or the slave, depending on the master's transmit mode or receive mode.

\begin{figure}[t]
	\centering
	\includegraphics[width=0.98\cw]{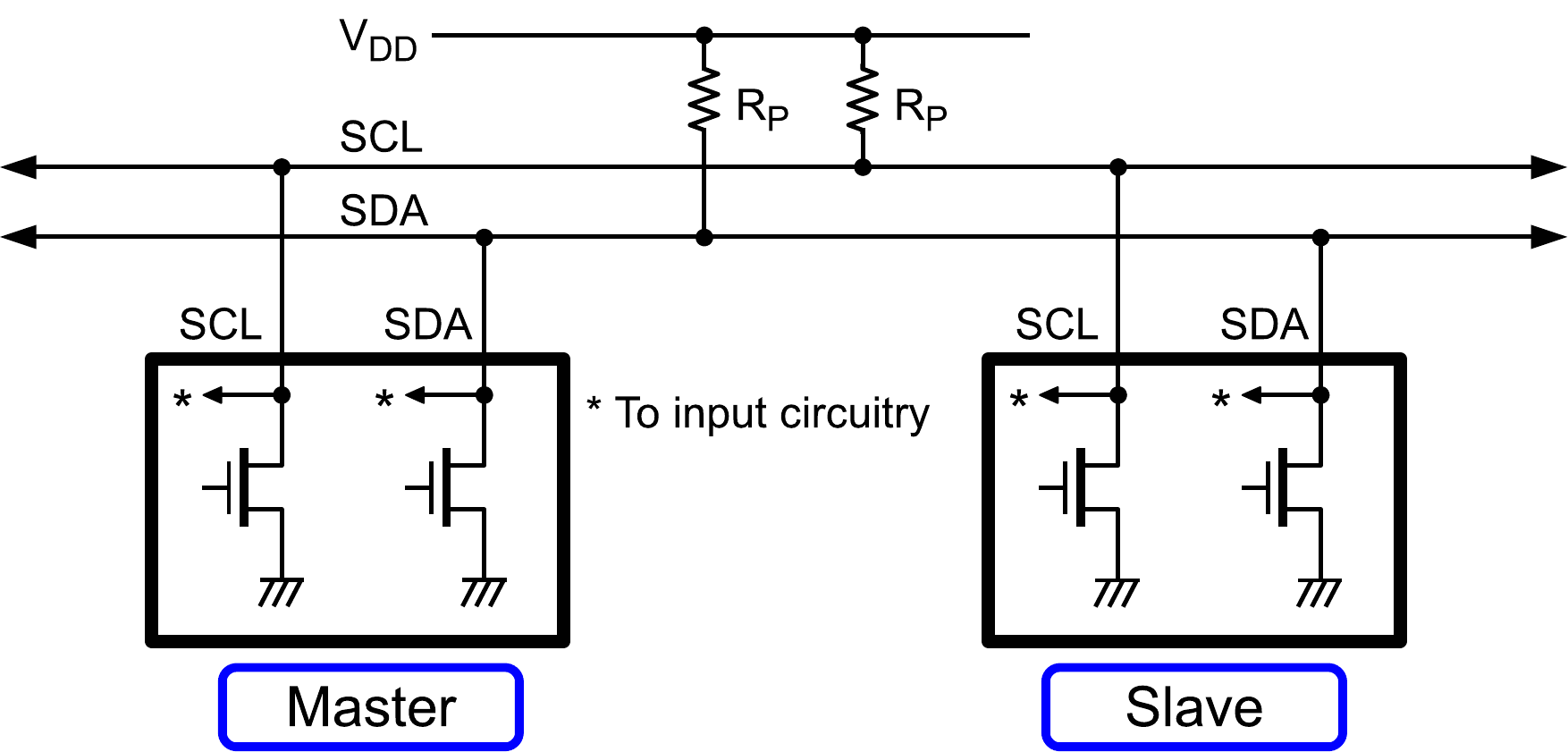}
	\caption{\iic\ bus configuration.
		{Digital input circuitries, omitted herein, 
		draw a negligible current from the bus. }}
	\label{fig:i2c}
\end{figure}

An \iic\ bus configuration including the output transistors inside the master and the slave is shown in \reffig{i2c}.
The SCL and SDA pins of each node are bidirectional and are connected internally to an input circuitry and an open-drain transistor output. 
{The digital input circuitry draws only a negligible current.}
The SCL and SDA lines are pulled up individually to the positive supply voltage with resistors. 
While no signal is transferred on the bus, all the output transistors are in the off state, and the SCL/SDA lines are nearly at the positive supply voltage, i.e., the logical high.
When any one of the output transistors is turned on, the signal line is inverted nearly to the ground level, i.e., the logical low.

It is noteworthy that the output pins only sink the current, but do not actively source the current. 
The dc voltage required to interpret the transistors' off/on states to logical high/low levels is supplied externally. 

\subsection{Clock and Data Multiplexing on Conductive Textile}

As described in \refsec{textile}, a single transmission line is formed with a double-sided conductive textile. 
In this work, the SCL/SDA signals and dc power supply are transferred simultaneously on the same transmission line using FDM.
For SCL/SDA transmission, two frequency carriers are modulated respectively with SCL and SDA, in the form of amplitude shift keying (ASK).

\begin{figure}[t]
	\centering
	\includegraphics[width=0.98\cw]{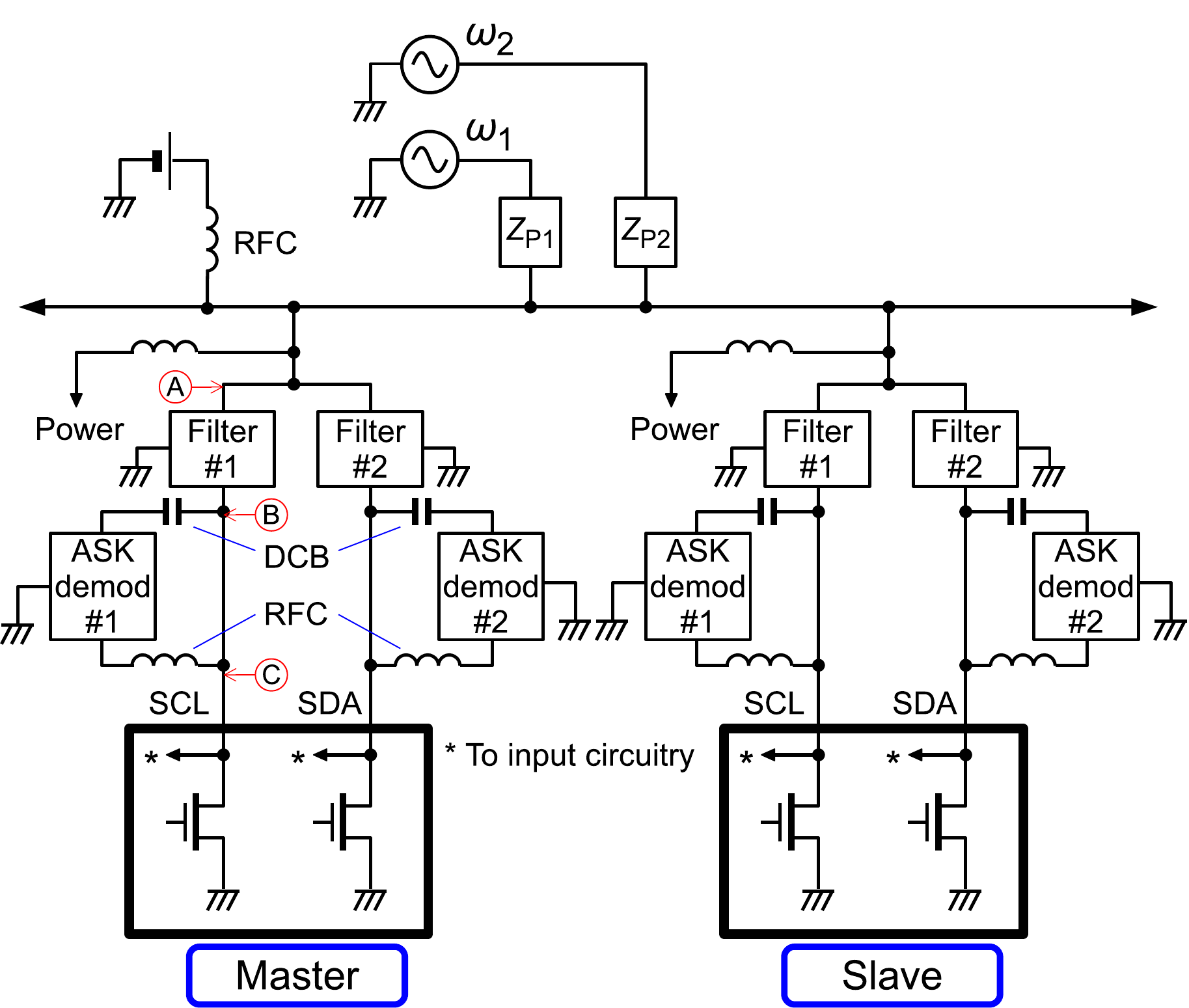}
	\caption{\iiw\ bus configuration.
		Labels A, B, and C are to indicate the corresponding points 
		in Figs. \ref{fig:passivemod}, \ref{fig:tnet}, \ref{fig:demodblock} 
		and \ref{fig:demodschem}. }
	\label{fig:i2w}
\end{figure}

The proposed \iiw\ bus configuration including the output transistors inside the master and the slave is shown in \reffig{i2w}.
Similar to the external dc supply to the signal lines via the pull-up resistors in conventional \iic, the RF carriers are supplied externally to the bus via ``pull-up'' impedances.
The carrier generator can be separated from the master and the slave.
The SCL and SDA pins of each node are connected to the bus through special bandpass filters, and the output transistors' off/on states are interpreted directly into high/low states in the RF carrier amplitude on the signal line. 
In each of the master and the slave, the amplitude shift is interpreted as logical high/low levels by the ASK demodulation circuitry.
Thus, the SCL/SDA pins of an off-the-shelf \iic-enabled ICs modulate the carriers directly and receive digital SCL/SDA signals via the demodulation circuitry. 

\subsection{Filter Design for Passive Modulation}
\label{ssec:filterdesign}

In this subsection, the ASK modulation operation and the filter design to achieve the operation are described in detail. 

The basic idea for passive ASK modulation in the configuration shown in \reffig{i2w} is, similar to the conventional \iic, to switch the impedance connected to the signal line between open and short, depending on the logical state of the SCL/SDA pin. 
To modulate two carriers independently, the impedance element should achieve an open/short switch at the frequency of the carrier to be modulated, and should remain open at the other frequency. 

\begin{figure}[t]
\begin{center}
	\includegraphics[width=0.95\cw]{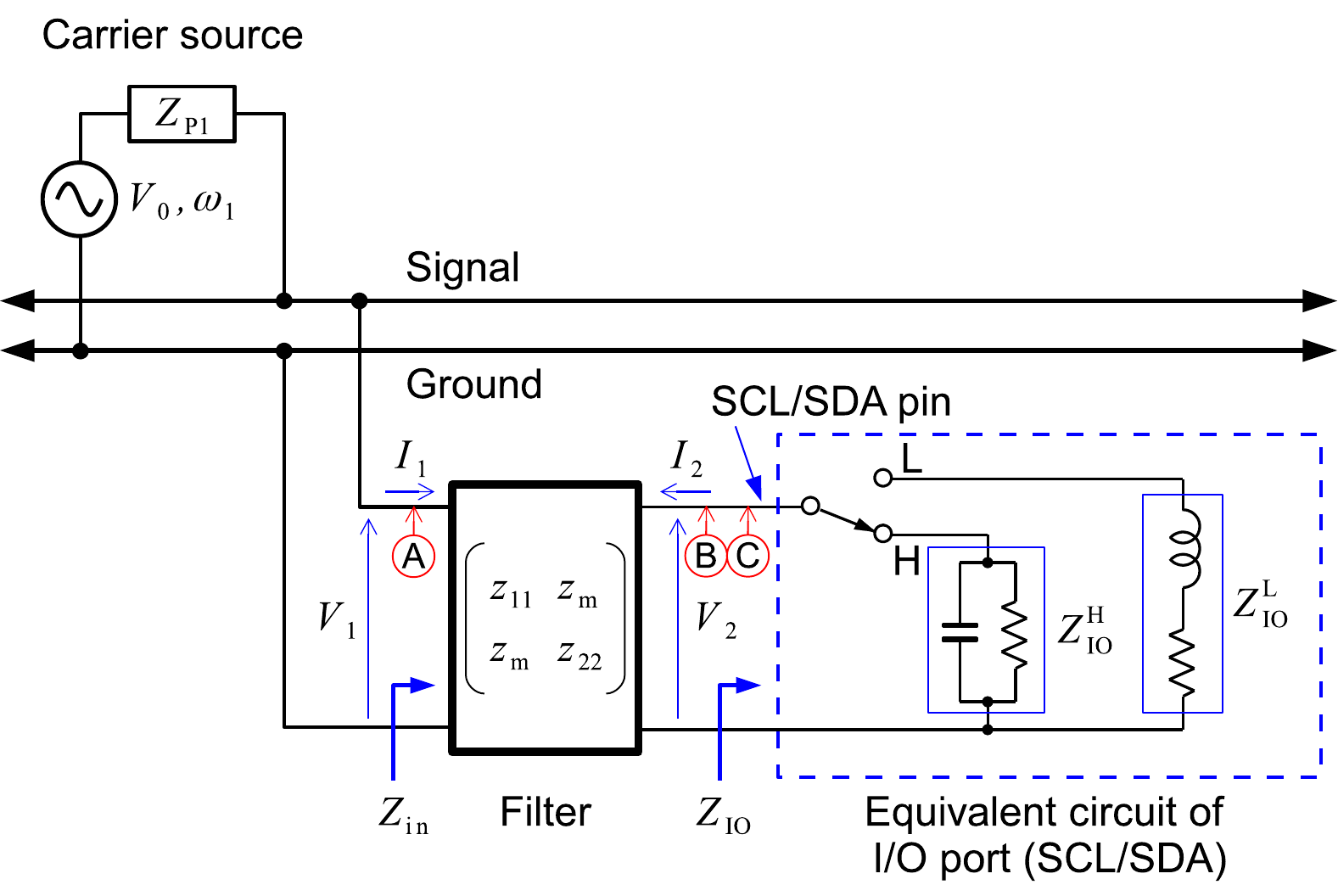}
\end{center}
\caption{Equivalent circuit for a filter and an SCL/SDA port. 
		Only one carrier source is shown. 
		Labels H and L at the switch terminals represent the logical high and low states, respectively. 
		At the logical high-state, the I/O is modeled as a parallel connection of 
		a small parasitic junction capacitance and a small conductance 
		representing a loss factor.
		At the logical low-state, it is modeled as a series network of 
		a small resistance and a small parasitic inductance.
		Labels A, B, and C correspond to those 
		in \reffig{i2w}.
		}
\label{fig:passivemod}
\end{figure}

An equivalent circuit that represents one of the two (SCL and SDA) signal inputs/outputs (I/Os), its associating filter, and a carrier source is illustrated in \reffig{passivemod}.
{It is noteworthy that the model includes the parasitic impedance associated with both the input circuitry and output transistor of the I/O. Although the input impedance at the logical high-state, $\Zio^\hi$, draws only a negligible current at the frequency of the digital signals, it is not negligible at the carrier frequencies. The capacitance is not neglected, but rather intentionally utilized for the filter resonance.}

We define the impedance matrix of the {reciprocal} two-port filter as follows:
{%
\begin{align}
	\label{eq:zmatrix}
	\begin{pmatrix}
	V_1 \\
	V_2
	\end{pmatrix}
	=
	\begin{pmatrix}
	z_{11} & \zm \\
	\zm & z_{22}
	\end{pmatrix}
	\begin{pmatrix}
	I_1 \\
	I_2
	\end{pmatrix},
\end{align}}%
where $V_1$ and $V_2$ are the voltages across ports 1 and 2, respectively; $I_1$ and $I_2$ are the currents flowing into ports 1 and 2, respectively. 

The secondary port is connected to the SCL/SDA pin. 
Let $\Zio$ denote the input impedance of the SCL/SDA pin; {subsequently 
\begin{align}
	\label{eq:v2}
	V_2
=
	- \Zio I_2 .
\end{align}%
From \refeq{zmatrix} and \refeq{v2},}
the impedance looking into the filter at the primary port, $Z_\in$, is calculated as follows: 
{%
\begin{align}
\label{eq:zin}
	Z_\in
=
	\frac{V_1}{I_1}
=
	z_{11} - \frac{\zm^2}{\Zio+z_{22}}
\end{align}%
}%
Thus, the filter�fs primary port impedance depends on $\Zio$, which is determined by the on/off-state of the open-drain transistor, i.e., the logical high/low output.
Let $\Zio^\hi$ and $\Zio^\lo$ represent the impedances of the I/O at the logical high and low states, respectively.
While the $\omega_1$ carrier with a constant voltage amplitude $V_0$ is supplied via a pull-up impedance $\Zp$, the voltage appearing at the primary port of the filter is calculated as follows:
\begin{align}
\label{eq:v1}
	V_1
=
	\frac{Z_\in}{\Zp + Z_\in}V_0
\end{align}

From \refeq{zin} and \refeq{v1}, $V_1$ is determined by $Z_\in$ depending on $\Zio$. 
Thus, the $\omega_1$ carrier is ASK modulated directly depending on the logical output of the SCL/SDA pin. 
The filter for modulating the $\omega_1$ carrier should maintain a high impedance at $\omega_2$, regardless of the logical pin state. 

The requirements on $Z_\in(\omega)$ to modulate the $\omega_\mod$ carrier and to avoid the cross-talk modulation of the $\omega_\stop$ carrier are as follows:
\begin{enumerate}
\item For the logical high-state, $Z_\in(\omega_\mod)$ should be open, to avoid a decrease in the joint impedance of all the filters connected to the same transmission line. 
\item For the logical low-state,  $Z_\in(\omega_\mod)$ should be short, to achieve a higher modulation depth for clearly distinguishable symbols. 
\item Regardless of the logical state, $Z_\in(\omega_\stop)$ should be open, to avoid cross-talk modulation.
\end{enumerate}
The requirements above are summarized in \reftab{zin}.
\begin{table}[b]
\caption{Ideal input impedance of filter, $|Z_\in(\omega)|$} 
\label{tab:zin}
\centering
\begin{tabular}{ccc}
\hline
& $\Zio=\Zio^\hi$ & $\Zio=\Zio^\lo$\\
\hline\hline
$\omega=\omega_\mod$ & $\infty$ $^{1)}$ & $0$ $^{2)}$ \\
$\omega=\omega_\stop$ & $\infty$ $^{3)}$ & $\infty$ $^{3)}$\\
\hline
\end{tabular}\\
{\footnotesize 1), 2), 3): Correspond to the enumerated items in the body text.}
\end{table}

Suppose that the filter consists of lossless, passive reactances. 
All the elements of the impedance matrix in \refeq{zmatrix} are therefore pure imaginary numbers.
Subsequently, the impedance matrix is realized with a T-network shown in \reffig{tnet}, where the reactance on each branch is defined as follows:
\begin{figure}[t]
	\centering
	\includegraphics[width=0.35\cw]{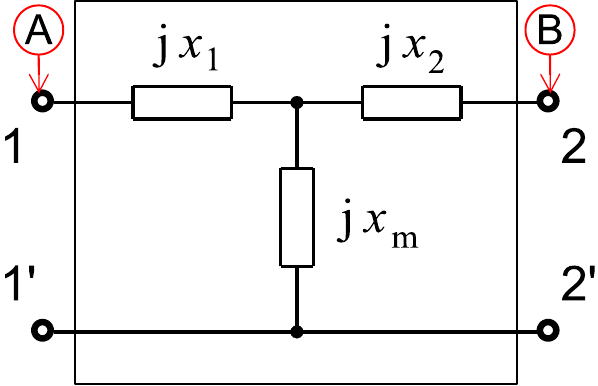}
	\caption{Realization of the impedance matrix in \refeq{zmatrix}.
	Labels A and B correspond to those in \reffig{i2w}.}
	\label{fig:tnet}
\end{figure}
\begin{align}
	x_1
	&=
	\im{z_{11}-\zm}
	\\
	\label{eq:x_2}
	x_2
	&=
	\im{z_{22}-\zm}
	\\
	\label{eq:xm}
	\xm
	&=
	\im{\zm}.
\end{align}
The three elements in the T-network are determined by considering the requirements, as described below.

Because the I/O pin is a parallel network of the digital input circuitry and the open-drain transistor in the off-state, $\Zio^\hi$ is an almost pure capacitive reactance, i.e.,
\begin{align}
	\Zio^\hi
\approx
	-\j \Xioh,
\end{align}
where $\Xioh$ is {a} positive real number.

The first requirement stated above, $Z_\in\rightarrow\infty$ for the logical high state, can be satisfied if $\zm \neq 0$ and $\Zio^\hi+z_{22}=0$ in \refeq{zin}.
Therefore, $z_{22}$ is expressed as follows:
\begin{align}
\label{eq:z22xio}
	z_{22}
=
	\j\Xioh,
\end{align}
which is a pure inductive reactance. 

\begin{figure}[t]
	\centering
	\includegraphics[width=0.7\cw]{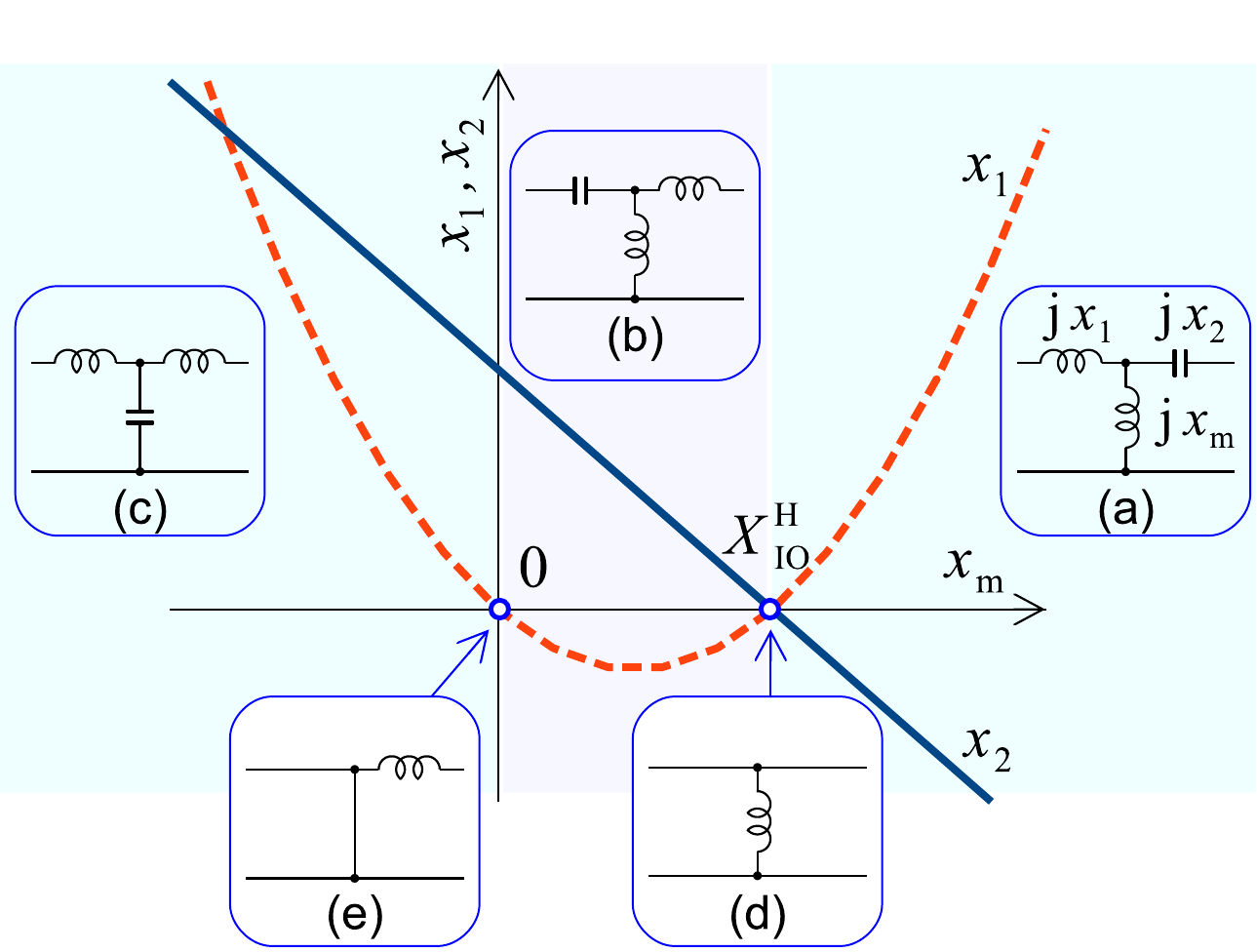}
	\caption{Combinations of inductive/capacitive reactances.
		Five configurations exist depending on $\xm$: 
		(a) $\xm>\Xioh$, (b) $0<\xm<\Xioh$, (c) $\xm<0$, (d) $\xm=\Xioh$, and (e) $\xm=0$.}
	\label{fig:reactance}
\end{figure}

The second requirement, $Z_\in=0$ for the logical low, is achieved by $z_{11} = {\zm^2}/({\Zio^\lo+z_{22}})$.
This can be approximated as $z_{11} \approx {\zm^2}/z_{22}$, because $|z_{22}|=|\Zio^\hi|\gg|\Zio^\lo|$.
Therefore, 
\begin{align}
\label{eq:z11z22zm}
	z_{11} - \zm
\approx
	\frac{\zm^2}{z_{22}} - \zm
=
	- \frac{\zm}{z_{22}}(z_{22}-\zm).
\end{align}
By substituting \refeq{z22xio}, \refeq{x_2} and \refeq{z11z22zm} are respectively rewritten as follows:
\begin{align}
\label{eq:x2}
	x_2
=
	\Xioh - \xm
\end{align}
and 
\begin{align}
\label{eq:x1}
	x_1
=
	- \frac{\xm}{\Xioh}x_2.
\end{align}
The relationship among $x_1, x_2$, and $\xm$, and combinations of the inductive/capacitive reactances satisfying the equations above are shown in \reffig{reactance}. 

For the third requirement, to stop the $\omega_\stop$ current from flowing through the filter�fs primary port regardless of the logical state of the I/O pin, an LC parallel resonant network should be placed at the $x_1$-branch. 
The joint impedance of a parallel connection of inductance $L_1$ and capacitance $C_1$, satisfying $\omega_\stop^2=(L_1 C_1)^{-1}$, at $\omega=\omega_\mod$ is calculated as follows.
\begin{align}
\label{eq:x1omegamod}
	\j x_1(\omega_\mod)
&=
	\frac 1{\j\omega_\mod C_1 + (\j\omega_\mod L_1)^{-1}}\nonumber\\
&=
	\j\frac{\alpha}{1-\alpha^2}\omega_\stop L_1
=
	-\j\frac{\alpha}{\alpha^2-1}\frac1{\omega_\stop C_1},
\end{align}
where $\alpha\equiv\omega_\mod/\omega_\stop$.
Thus, $x_1(\omega_\mod)$ is inductive for $\omega_\mod<\omega_\stop$, and $x_1(\omega_\mod)$ is capacitive for $\omega_\mod>\omega_\stop$.
An appropriate configuration should be chosen depending on $\alpha$.
For example, $x_1$ is inductive in configuration (a) shown in \reffig{reactance}; therefore, this configuration is applicable only to the condition $\omega_\mod<\omega_\stop$. 
Meanwhile, for $\omega_\mod>\omega_\stop$, configuration (b) is applicable.

To summarize the discussions above, the minimal filter configurations are listed in \reftab{filterconf}.
\begin{table*}[b]
\caption{Minimal Filter Configurations {Corresponding to \reffig{reactance}.}}
	\centering
	\begin{tabularx}{\linewidth}{Xccccc}
\hline
&(a)&(b)&(c)&(d)-1&(d)-2 \tabularnewline
\hline
Frequency condition &$\omega_\stop>\omega_\mod$&$\omega_\stop<\omega_\mod$&$\omega_\stop>\omega_\mod$&$\omega_\stop>\omega_\mod$ &$\omega_\stop<\omega_\mod$
\vspace{3mm}
\tabularnewline
Configuration&
\begin{minipage}{26mm}
\centering
\scalebox{0.6}{\includegraphics{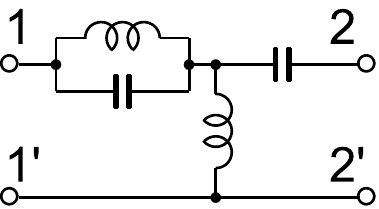}}
\end{minipage} &
\begin{minipage}{26mm}
\centering
\scalebox{0.6}{\includegraphics{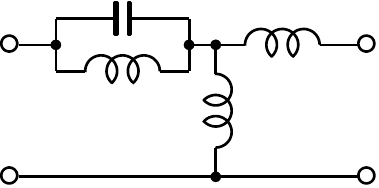}}
\end{minipage} &
\begin{minipage}{26mm}
\centering
\scalebox{0.6}{\includegraphics{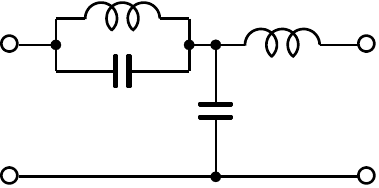}}
\end{minipage} &
\begin{minipage}{26mm}
\centering
\scalebox{0.6}{\includegraphics{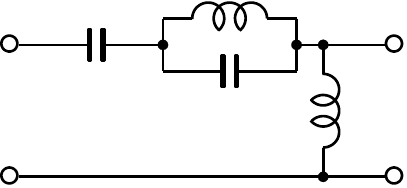}}
\end{minipage} &
\begin{minipage}{26mm}
\centering
\scalebox{0.6}{\includegraphics{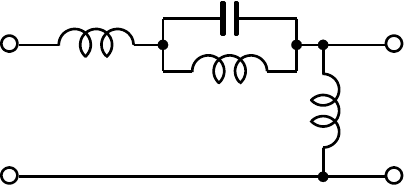}}
\end{minipage} 
\vspace{3mm}
\tabularnewline
Equivalent circuit\newline at $\omega=\omega_\mod$&
\begin{minipage}{26mm}
\centering
\scalebox{0.6}{\includegraphics{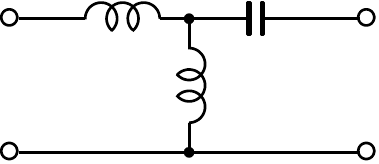}}
\end{minipage} &
\begin{minipage}{26mm}
\centering
\scalebox{0.6}{\includegraphics{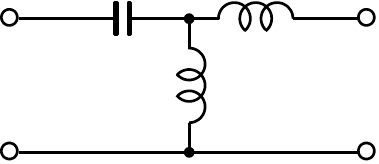}}
\end{minipage} &
\begin{minipage}{26mm}
\centering
\scalebox{0.6}{\includegraphics{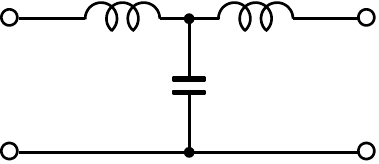}}
\end{minipage} &
\begin{minipage}{26mm}
\centering
\scalebox{0.6}{\includegraphics{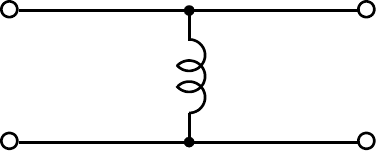}}
\end{minipage} &
\begin{minipage}{26mm}
\centering
\scalebox{0.6}{\includegraphics{dmod.pdf}}
\end{minipage}
\vspace{3mm}
\tabularnewline
Equivalent circuit\newline at $\omega=\omega_\stop$&
\begin{minipage}{26mm}
\centering
\scalebox{0.6}{\includegraphics{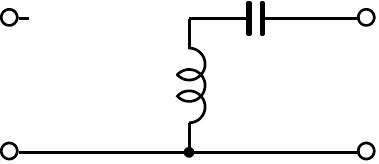}}
\end{minipage} &
\begin{minipage}{26mm}
\centering
\scalebox{0.6}{\includegraphics{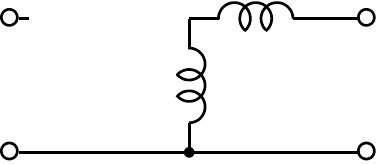}}
\end{minipage} &
\begin{minipage}{26mm}
\centering
\scalebox{0.6}{\includegraphics{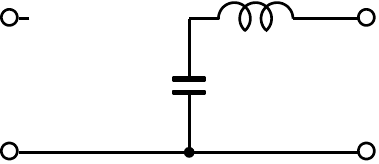}}
\end{minipage} &
\begin{minipage}{26mm}
\centering
\scalebox{0.6}{\includegraphics{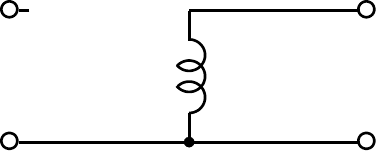}}
\end{minipage} &
\begin{minipage}{26mm}
\centering
\scalebox{0.6}{\includegraphics{dstop.pdf}}
\end{minipage}
\vspace{3mm}
\tabularnewline
\hline
\end{tabularx}
\label{tab:filterconf}
\end{table*}
(a), (b), (c), and (d) on the top row of the table correspond to those in \reffig{reactance}.
Configuration (d), in which $x_1(\omega_\mod)=x_2(\omega_\mod)=0$, is realized by series resonance at the $x_1$-branch. 
The additional reactance element connected in series to the parallel LC is inductive or capacitive, depending on $\alpha$. 
Configuration (e) shown in \reffig{reactance} is inappropriate and is therefore excluded from the table, because all of the current flowing into the filter from the primary port flows through $\xm=0$ and no current appears at the secondary port.

\subsection{ASK Demodulator}
\label{ssec:demod}

To interpret the carrier amplitude shift into a logical high/low that is comprehensive for the digital IC, an ASK demodulator is required for each SCL/SDA pin, as shown in \reffig{i2w}. 

The demodulator input is ac-coupled with a capacitor, and the output is RF-decoupled with an inductor. 
Therefore, the demodulator operates without any positive/negative feedback effects such as oscillation and self-holding, although the input and output are connected to the same SCL/SDA pin.

\begin{figure}[t]
	\centering
	\includegraphics[width=0.75\cw]{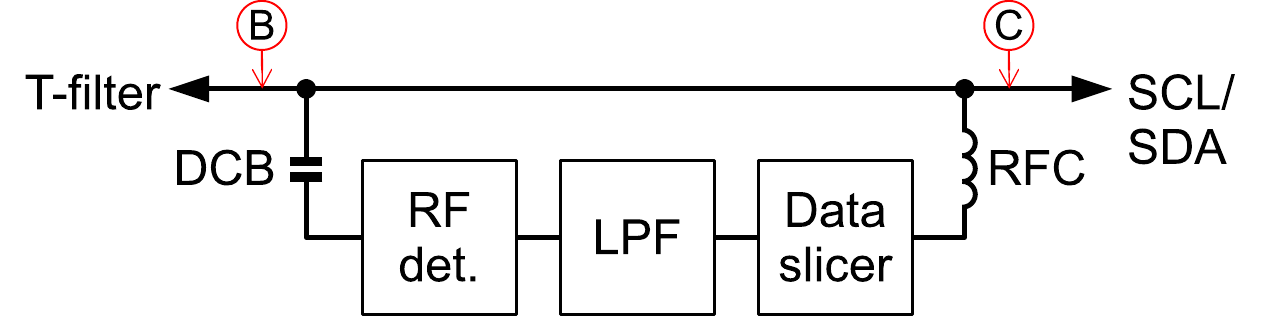}
	\caption{ASK demodulator.
	Labels B and C correspond to those in \reffig{i2w}.}
	\label{fig:demodblock}
\end{figure}

A block diagram of the demodulator is shown in \reffig{demodblock}.
The $\omega_\mod$ carrier is fed into the RF detector via the filter described in the previous subsection, and the $\omega_\stop$ carrier is blocked by the filter. 
The RF detector extracts the envelope of the $\omega_\mod$ carrier.
The analog envelope waveform is binarized by a data slicer \cite{dataslicer}. 
The data slicer tunes the reference voltage of the comparator automatically depending on the input waveform, via a lowpass filter (LPF). 
The binary output of the data slicer is accepted by the SCL/SDA pin.
The comparator output terminal should be RF-decoupled from the filter and the SCL/SDA pin using an RF choke (RFC), such that the comparator output impedance does not affect the filter design.

\section{Design Example}
\label{sec:example}

This section presents an example of the \iiw\ system design. 
It is noteworthy that it is only an example, and is not an optimized design. 

\subsection{Carriers and I/O Capacitance}

To design the filters, the carrier frequencies and the capacitance of the SCL/SDA pins must be specified. 

We choose the carrier frequencies of 20 MHz and 50 MHz for the SCL and SDA, respectively. 
The two frequencies should avoid being an integer multiple of each other, to minimize crosstalk due to harmonics. 
The lower bound of the carrier depends on the frequency of the SCL signal. 
For the standard mode with 100-kHz SCL, the carrier frequencies are required to be significantly higher than 100 kHz.
The upper bound is determined such that the wavelength in the textile medium is significantly longer than the size of the textile. 
If the wavelength is comparable to or shorter than the size of the textile, the carrier signal strength at each receiver depends primarily on the location and frequency, because of the standing waves. 
The received signal strength fluctuation requires a more complicated configuration of the demodulator.

In this work, we used a development board called Teensy 3.2 \cite{teensy32} with a 32-bit MCU, MK20DX256VLH7 \cite{k20}. 
The measured capacitance of the SCL/SDA pin of the MCU, including the stray capacitance of PCB copper tracks, was approximately 8 pF at the logical high state.
For the carrier frequencies 20 MHz and 50 MHz, the reactance of the capacitance is approximately $-\j1$ k$\Omega$ and $-\j400$ $\Omega$, respectively. 

\subsection{Filter Design}

{As discussed in \refssec{filterdesign},}
The filter configurations should be selected from those shown in \reftab{filterconf}.
Here, we choose (a) and (b) for the lower carrier (20 MHz) and higher carrier (50 MHz), respectively. 

The theoretical analyses of the filter presented in the previous section assumed lossless reactance components. 
A one degree of freedom exists and $\xm$ can be determined arbitrarily to satisfy $\xm>\Xioh$ in (a) and $0<\xm<\Xioh$ in (b). 
For the actual filter design, the one degree of freedom enables the appropriate inductors and capacitors to be chosen such that the filter possesses a significantly high/low impedance to achieve the characteristics shown in \reftab{zin}.
The highest/lowest impedance that can be achieved by the filter depends on the specific characteristics of the inductors and capacitors, including the loss factors.

One significant factor that dominates the filter characteristics is the frequency of the impedance zero between two impedance poles.
When the SCL/SDA pin is at the logical high state, two poles of $Z_\in(\omega)$ are required to be located at $\omega=\omega_\mod$ and $\omega=\omega_\stop$.
From Foster's reactance theorem \cite{foster1924}, there is an impedance zero at $\omega=\omega_0$, where $\omega_\mod<\omega_0<\omega_\stop$ for configuration (a) and $\omega_\stop<\omega_0<\omega_\mod$ for (b). 
The zero should be separated from the two poles on the frequency axis, because a pole can be almost canceled by the closely located zero, and the impedance magnitude will be reduced significantly .
The frequency of the impedance zero can be tuned with $\xm$.

The filter configurations including the equivalent impedance of the SCL/SDA pins are shown in \reftab{example}.
An example of the selection of inductors and capacitors is shown in \reftab{components}; it includes the exact values calculated with \refeq{x2} and \refeq{x1} and approximated values selected from the E12 series.
$\Cioh$ is increased by adding an external shunt capacitor in configuration (a) to reduce the inductance of $\xm>\Xioh$ such that the inductor can be chosen from commercially available high-Q wire-wound tiny chip inductors, which are typically lower than 10 $\upmu$H at the maximum.
Terminals 1-1', 2-2', and 1-2 are required to be isolated at dc; therefore, dc block (DCB) capacitors are inserted as shown in \reftab{example}. 
The impedance of the DCBs at the carrier frequencies are negligible.

\begin{figure}[t]
\begin{center}
	\subfloat[]{\resizebox{0.99\cw}{!}{\input{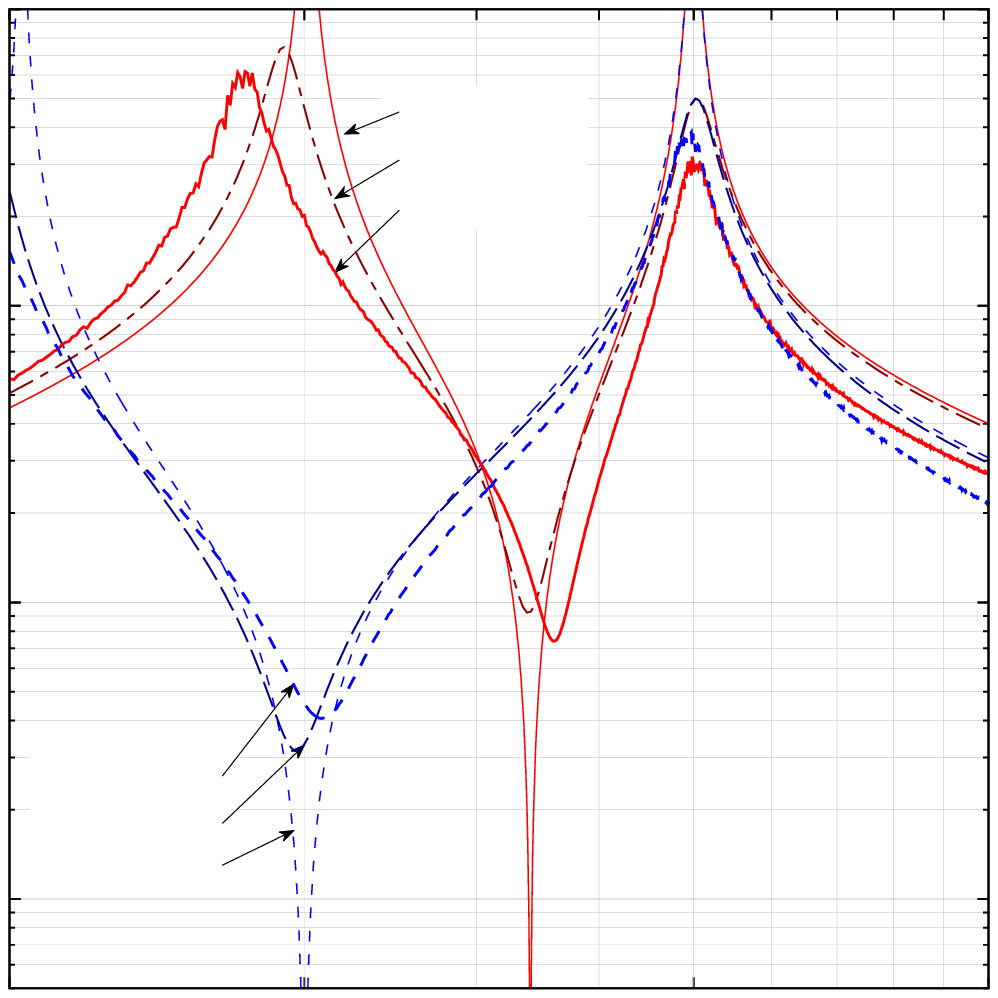}}}
	\\
	\subfloat[]{\resizebox{0.99\cw}{!}{\input{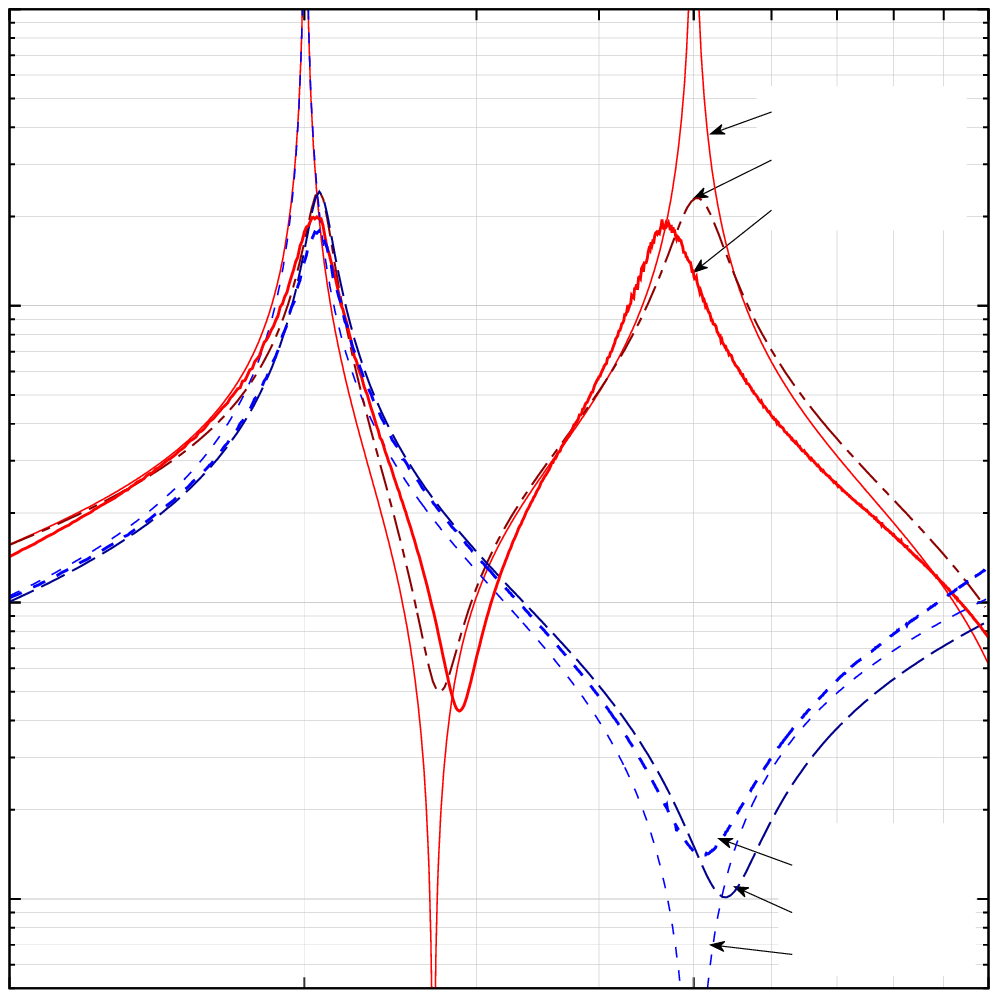}}}
\end{center}
\caption{Simulated filter input impedance of 
	configurations (a) and (b).
	The impedance values were simulated at both the logical high (H) and low (L) states 
	for each of the lossless and lossy models.
	Filters consisting of lossless components with exact values calculated from \refeq{x2} and 
	\refeq{x1} become either ideal open or short at the carrier frequencies.
	The lossy models also achieve significant impedance changes by two orders of magnitude.
	{Measured impedances of the fabricated filters shown in \reffig{filterphoto} are also plotted. }	
	}
\label{fig:zinfreq}
\end{figure}

The simulated impedance--frequency curves of the designed filters are presented in \reffig{zinfreq}. 
The simulations were performed with LTspice XVII. 
The simulation results for both lossless and lossy models are plotted. 
The lossless simulations were conducted with the exact calculated inductances/capacitances in \reftab{components}.
Additionally, $\Rioh$ was replaced with an open circuit and $\Riol$ was replaced with a short circuit. 
For lossy simulations, we used the E12 series values and manufacturer-supplied inductor SPICE models \cite{epcosEn} that represent the realistic characteristics well, including the losses and self-resonance.
The capacitors were modeled as ideal capacitances because their loss factors are negligible compared with the inductors at the frequencies of interest. 
Based on our preliminary measurement results, we chose $\Rioh=10$ k$\Omega$ and $\Riol=10$ $\Omega$ for the lossy models.
The result indicates that the lossless models achieve the ideal open and short at the carrier frequencies, and that the lossy models achieve significant impedance changes of more than two orders of magnitude at $\omega=\omega_\mod$.

\begin{table*}[b]
\caption{Filter design example}
	\centering
	\begin{tabularx}{0.65\tw}{Xcc}
\hline
&(a)&(b) \tabularnewline
\hline
Frequency condition &$\omega_\stop>\omega_\mod$&$\omega_\stop<\omega_\mod$
\vspace{3mm}
\tabularnewline
Configuration&
\begin{minipage}{40mm}
\centering
\scalebox{0.8}{\includegraphics{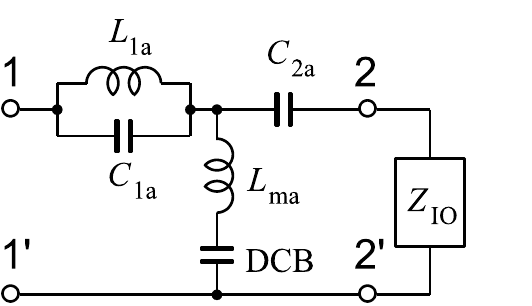}}
\end{minipage} &
\begin{minipage}{40mm}
\centering
\scalebox{0.8}{\includegraphics{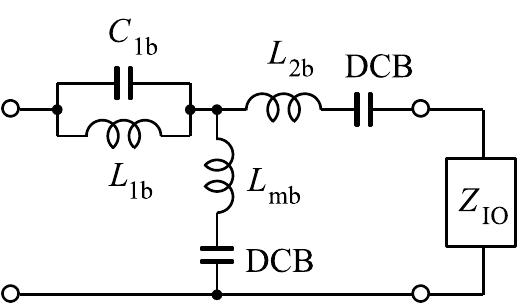}}
\end{minipage}
\vspace{3mm}
\tabularnewline
Equivalent circuit\newline at $\omega=\omega_\mod$\newline for logical high&
\begin{minipage}{40mm}
\centering
\scalebox{0.8}{\includegraphics{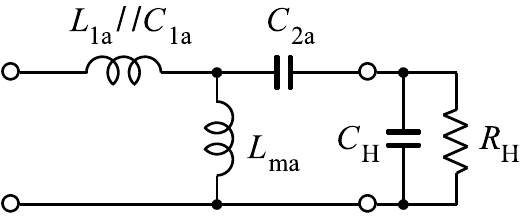}}
\end{minipage} &
\begin{minipage}{40mm}
\centering
\scalebox{0.8}{\includegraphics{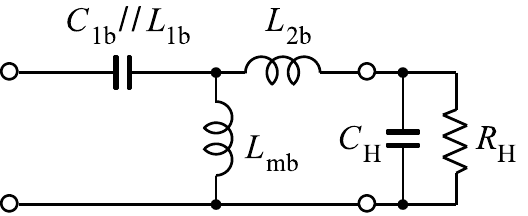}}
\end{minipage}
\vspace{3mm}
\tabularnewline
Equivalent circuit\newline at $\omega=\omega_\mod$\newline for logical low&
\begin{minipage}{40mm}
\centering
\scalebox{0.8}{\includegraphics{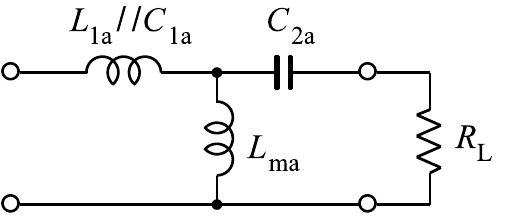}}
\end{minipage} &
\begin{minipage}{40mm}
\centering
\scalebox{0.8}{\includegraphics{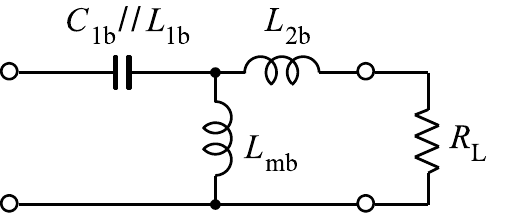}}
\end{minipage}
\vspace{3mm}
\tabularnewline
\hline
\end{tabularx}\\
{\footnotesize $L/\!/C$ represents parallel connection of $L$ and $C$.}
\label{tab:example}
\end{table*}

\begin{table}[b]
\caption{Inductors and capacitors} 
\label{tab:components}
\centering
\begin{tabular}{ccc|ccc}
\hline
\multicolumn{3}{c|}{(a)} & \multicolumn{3}{c}{(b)} \\
& Calculated & E12 series & & Calculated & E12 series \\
\hline\hline
$\Cioh$ & \multicolumn{2}{c|}{18 pF (8 pF $+$ 10 pF$^*$)} & $\Cioh$ & \multicolumn{2}{c}{8 pF}\\
$\Lma$ & \multicolumn{2}{c|}{ 4.7 $\upmu$H} & $\Lmb$ & \multicolumn{2}{c}{ 1.0 $\upmu$H} \\
$\La$ & 1.33 $\upmu$H & 1.2 $\upmu$H & $\Lb$ & 1.10 $\upmu$H & 1.0 $\upmu$H \\
$\Ca$ & 7.64 pF  & 8 pF & $\Cb$ & 57.3 pF & 56 pF\\
$\CCa$ & 53.6 pF  & 56 pF & $\LLb$ & 0.267 $\upmu$H & 0.22 $\upmu$H \\
\hline
\end{tabular}\\
{\footnotesize * An external shunt capacitor is added.}
\end{table}

\begin{figure}[t]
	\centering
	\begin{overpic}[width=0.7\cw]{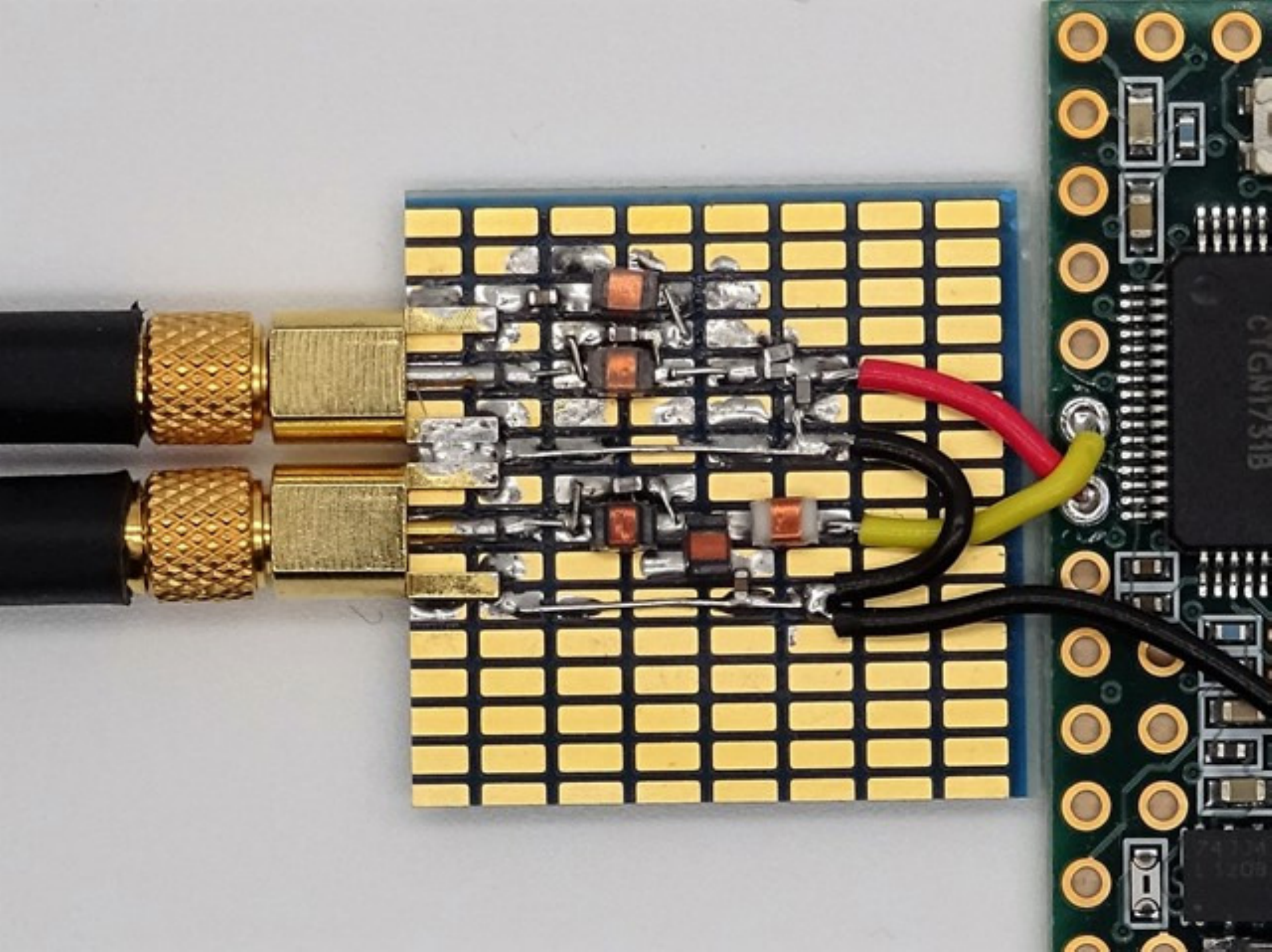}
		\put(0,0){\includegraphics[width=0.7\cw]{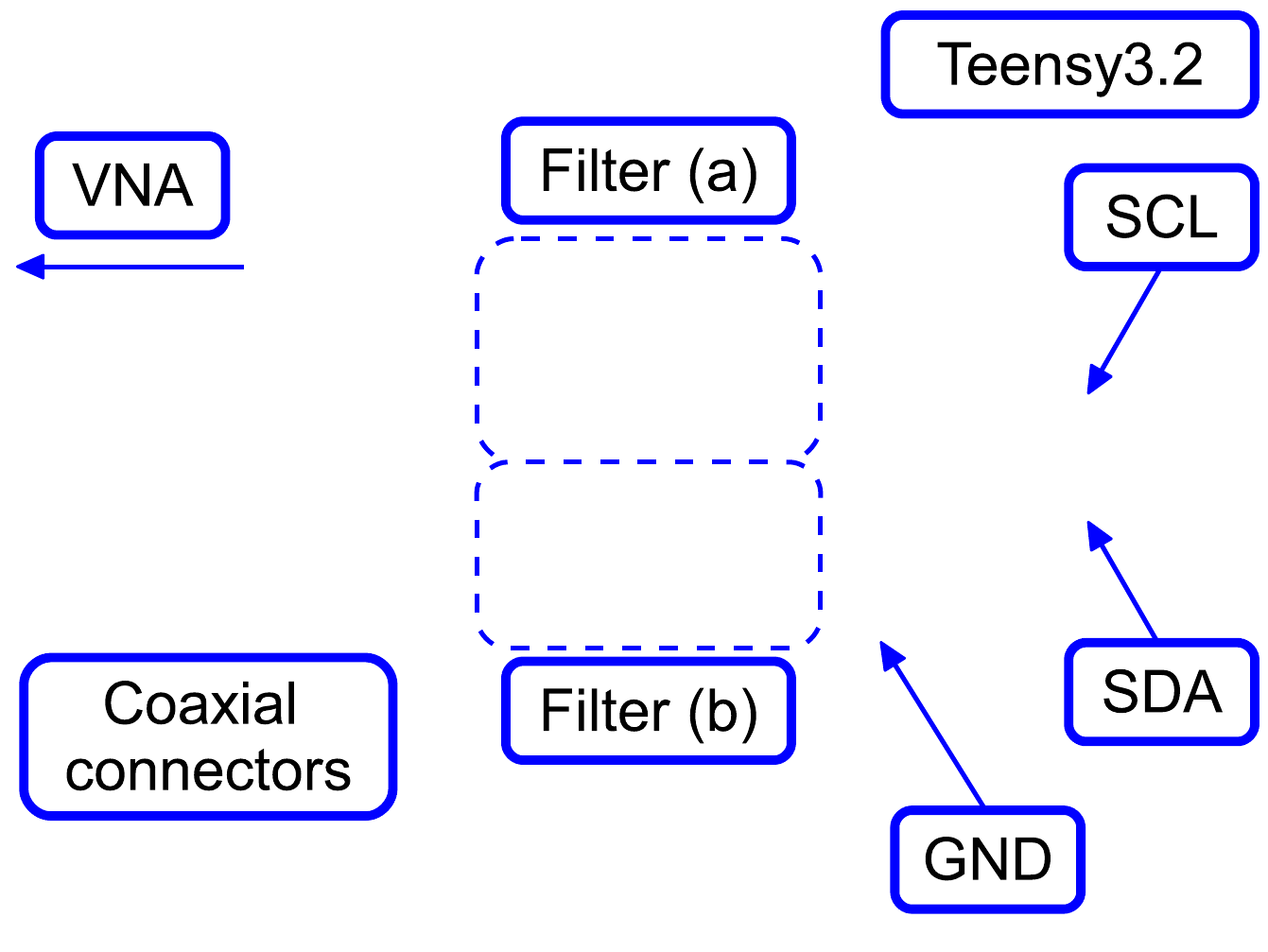}}
	\end{overpic}
	\caption{Impedance measurement of fabricated T-filters.}
	\label{fig:filterphoto}
\end{figure}

The filters (a) and (b) were fabricated as shown in \reffig{filterphoto}, and their input impedances were measured with a vector network analyzer (VNA) of Agilent Technologies E5071B.
The measurement results shown in \reffig{zinfreq} agree reasonably well with the simulation results of the lossy models. 
Let $\Zinh$ and $\Zinl$ denote $Z_\in$ at the logical high and low states, respectively. 
$\Zinh/\Zinl$ is approximately 44 at 20 MHz for filter (a), and 87 at 50 MHz for (b). 

In \reffig{passivemod}, let $\Vh$ and $\Vl$ denote the voltage across the primary port of the T-filter at the logical high and low states, respectively. 
The ratio $\Zinh/\Zinl$ determines the maximum ratios of $\Vh$ to $\Vl$. 
When the carrier is supplied from a voltage source via a pull-up impedance $\Zp$, $\Vh/\Vl$ is calculated {from \refeq{v1}} as follows.
\begin{align}
	\frac{\Vh}{\Vl}
=
	\frac{\Zinh/(\Zp+\Zinh)}{\Zinl/(\Zp+\Zinl)}
\end{align}
If $\Zp$ is significantly greater than $\Zinh$ and {$\Zinl$}, 
\begin{align}
\label{eq:ratio}
	\frac{\Vh}{\Vl}
\approx
	\frac{\Zinh}{\Zinl}.
\end{align}
Thus, the ratio \refeq{ratio} can achieve 44 and 87 at the maximum, for the SCL and SDA, respectively. 
However, the supplied carrier voltage attenuates significantly  with $\Zp\gg\Zinh$. 
$\Zp$ must be tuned according to the carrier source voltage and RF detector sensitivity.

{%
The ratio $\Vh/\Vl$ that determines the ASK modulation depth can be represented by \refeq{ratio} provided that only a single set of modulation circuitry is connected to the bus. 
When $n$ modules are connected in parallel and $n-1$ modules maintain a high-impedance state during the modulation operation of a single module, \refeq{ratio} is revised as follows: 
\begin{align}
\label{eq:ratiorev}
	\frac{\Vh}{\Vl}
\approx
	\frac{\Zinh/n}{\Zinl/\!/\{\Zinh/(n-1)\}},
\end{align}
where $/\!/$ represents parallel connection.
For $n\gg 1$,
\begin{align}
\label{eq:ratiorev2}
	\frac{\Vh}{\Vl}
\approx
	1+\frac{\Zinh}{n\Zinl}.
\end{align}
For $n \geq \Zinh/\Zinl$, $\Vh/\Vl \leq 2 \approx 6$ dB. 
Hence, the maximum number of the modules that can operate on the same bus depends on $\Zinh/\Zinl$ and the demodulator characteristics.}

\subsection{ASK Demodulator Design}

\begin{figure}[t]
	\centering
	\includegraphics[width=0.98\cw]{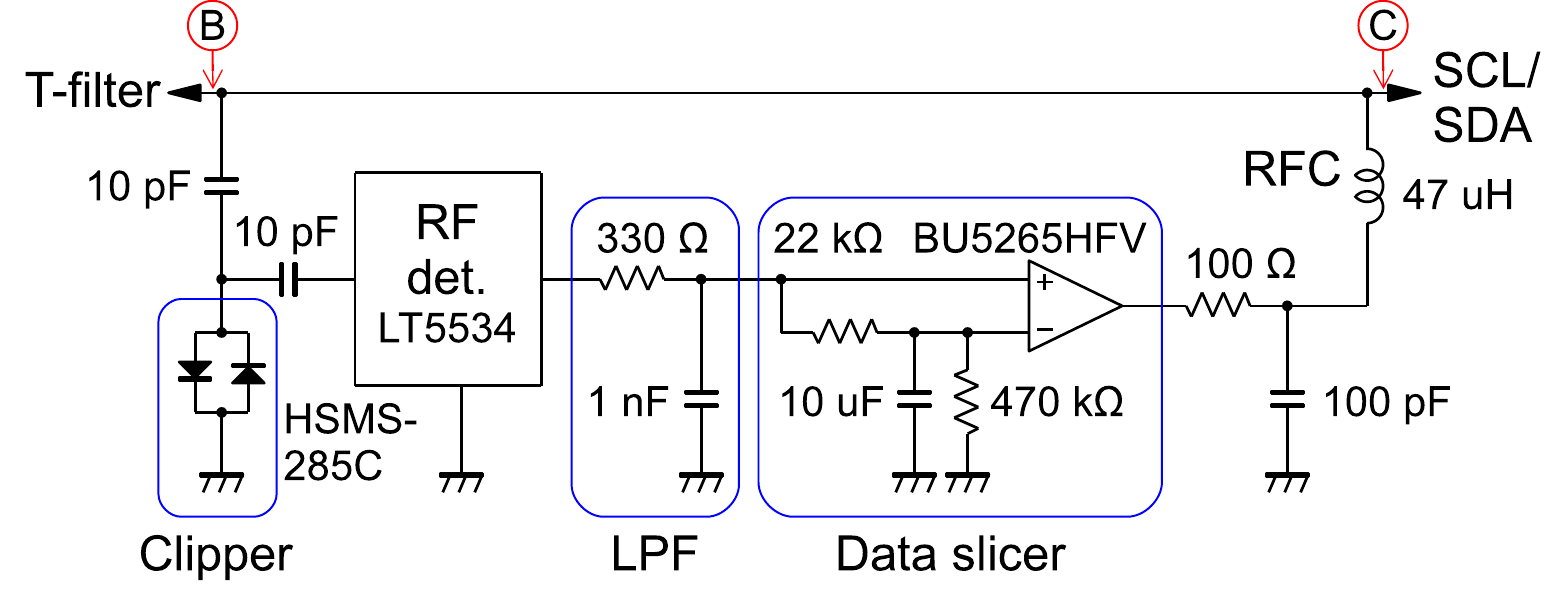}
	\caption{Design example of the ASK demodulator.
	Labels B and C correspond to those in \reffig{i2w}.
	The demodulator input and output are not directly shorted 
	at any frequencies,
	because the RF detector input is dc (baseband)-decoupled and the comparator output
	is RF-decoupled. }
	\label{fig:demodschem}
\end{figure}

A design example of the ASK demodulator is shown in \reffig{demodschem}.
The secondary port of the T-network filter is connected to the input of an RF detector via DCB capacitors. 
The envelope of the RF signal is detected and is fed to a data slicer via a low pass filter (LPF).

Although the DCB and RFC suppress feedback from the slicer output to the detector input as explained in \refssec{demod}, a transient voltage spike is induced at the detector input when the logic level of the SCL/SDA pin is toggled and stimulates the RF detector. 
This can cause the latch-up of the data slicer output as follows. 
When the RF amplitude shifts to the low level, the RF detector output decreases, and the data slicer output turns to low. 
The voltage step from the high to low level causes the  voltage spike at the RF detector input described above; subsequently, the detector output increases again, and the data slicer output becomes high. 
Thus, the slicer output holds the logical high, regardless of the actual RF amplitude.
To prevent the latch-up fault, an anti-parallel diode pair is connected to the input of the detector. The diodes clip off the spike at $\pm\Vf$, where $\Vf$ denotes the forward voltage drop of the diode. 
The desired RF signals $\omega_1$ and $\omega_2$ are detected accurately across the diodes, because the voltage amplitude of the RF signal is smaller than $\Vf$.

For the RF detector, LT5534 of Linear Technology was used. 
It is a log-scaled RF detector with the output slope of 44~mV/dB at 50 MHz and the input impedance of 2 k$\Omega$ \cite{lt5534}. 
The high input impedance is desirable to reduce the change in the impedance that is connected to the secondary port of the T-filter. 
A relatively small DCB capacitance that is a series connection of two 10-pF capacitors was used to further increase the impedance. 

The data slicer can be composed with general comparator ICs that operate with a bandwidth that is significantly greater than the SCL frequency, typically 100 kHz. 
In this example, Rohm BU5265HFV \cite{bu5265} was used. 
The resistor at the data slicer output is for limiting the shoot-through current. 
Although the slicer output and the SCL/SDA pin hold the same logical level, they can cause a discrepancy for a short period at a transition of the logical state and cause a shoot-through current. 
The capacitor connected to the resistor and the RFC is for the stabilization of the RF impedance looking into the demodulator from the output terminal, regardless of the logical state of the output. 

\begin{figure}[t]
	\centering
	\begin{overpic}[width=0.7\cw]{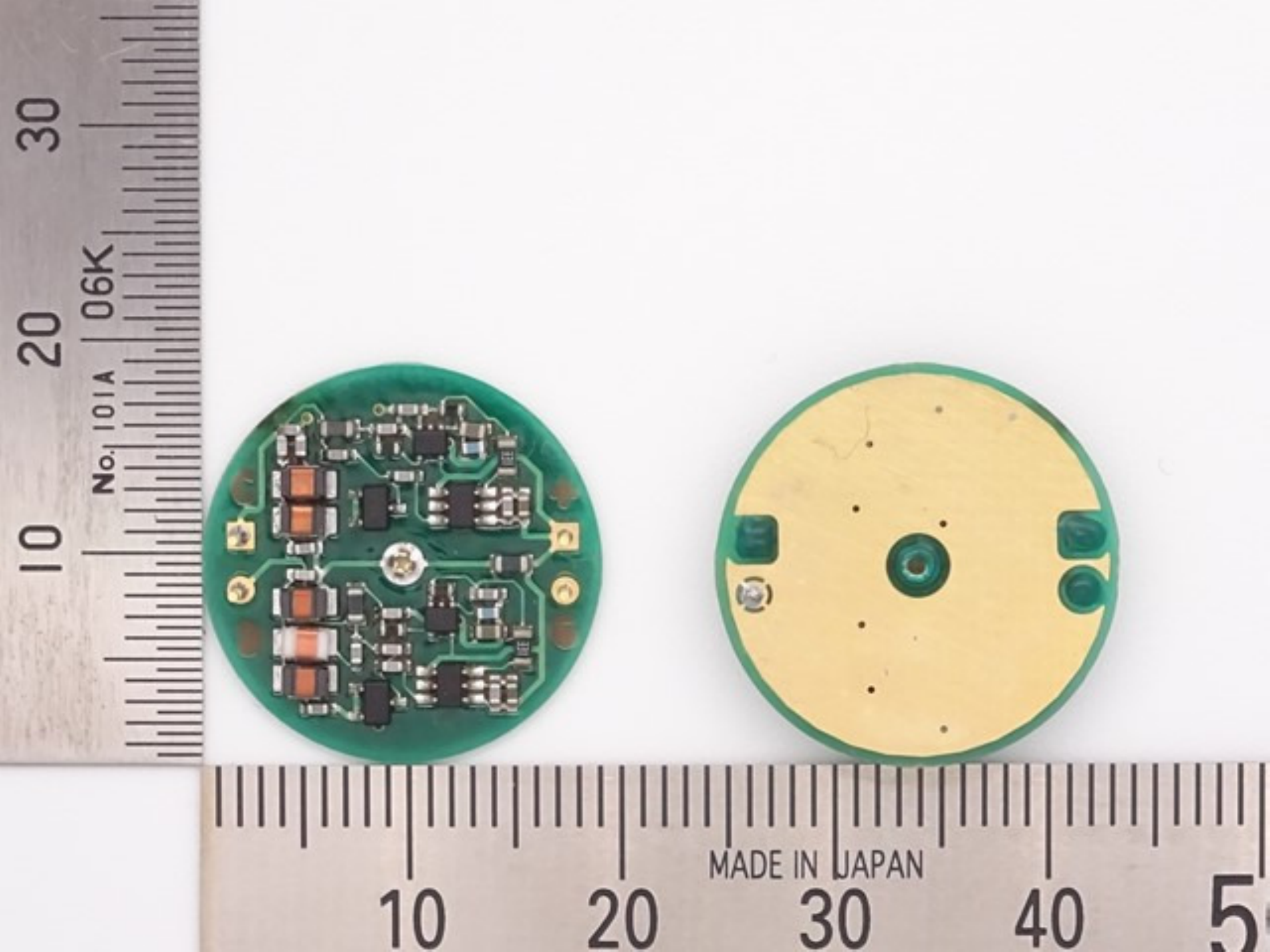}
		\put(0,0){\includegraphics[width=0.7\cw]{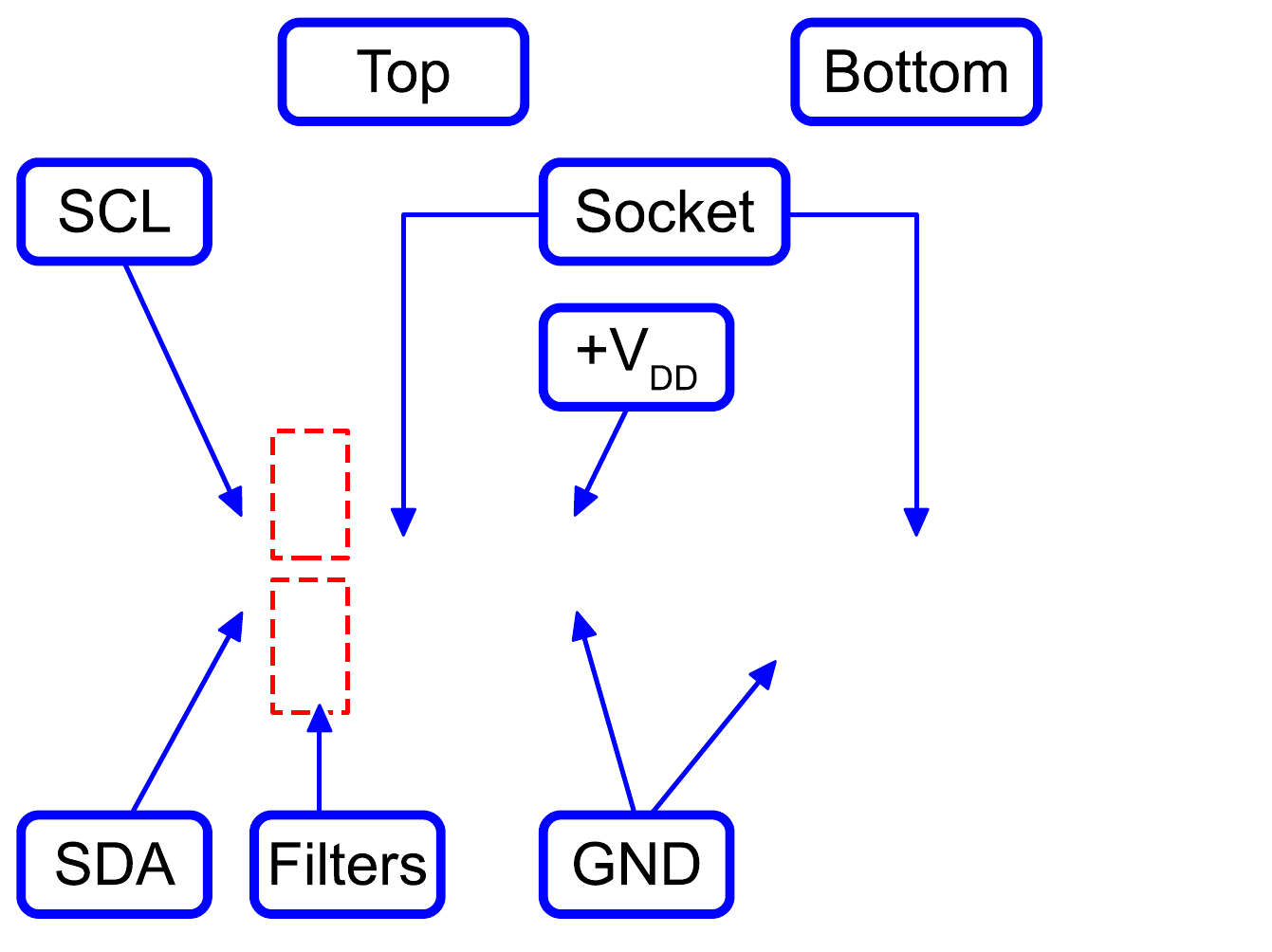}}
	\end{overpic}
	\caption{Fabricated modem module PCB.  
	}
	\label{fig:modemboard}
\end{figure}

\begin{figure}[t]
	\centering
	\resizebox{0.99\cw}{!}{\input{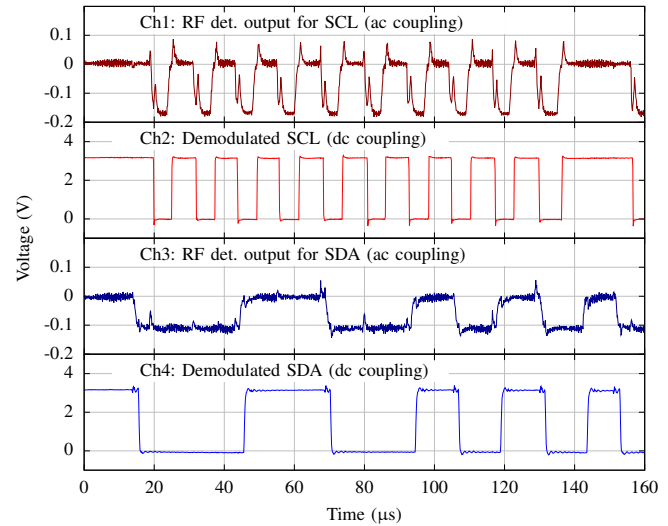}}
	\caption{Waveforms measured on the fabricated modem module.
	{	Ch1 and Ch3 channels are lowpass-filtered RF detector outputs measured at the positive input terminals of the comparators of SCL and SDA demodulators, respectively. 
	Ch2 and Ch4 are the demodulated waveforms measured at the SCL and SDA pins, respectively. 
	 Spikes are induced on Ch1 and Ch3 signals when the SCL logic level is toggled.
	 Nonetheless, their magnitudes are within the acceptable level for a stable demodulation.}}
	\label{fig:waveform}
\end{figure}

A modem module PCB was fabricated as shown in \reffig{modemboard}. 
A pair of filters and a pair of demodulators for the SCL and SDA were integrated into a single PCB. 
The measured waveforms on this module are shown in \reffig{waveform}.
In the measurement, 20-MHz and 50-MHz carriers of approximately 10--20 mV amplitude were supplied to the textile, and they were modulated with another modem module connected to the SCL/SDA ports of a Teensy 3.2 MCU board. 
{%
To clearly distinguish the baseband and carrier signals at the demodulator, their frequencies should be apart from each other by at least 1--2 orders of magnitude.}
For the power supply, a 3.3-V dc voltage was also applied to the textile via an RFC inductor. 

\section{Demonstration System}
\label{sec:demo}

\begin{figure}[t]
	\centering
	\begin{overpic}[width=0.98\cw]{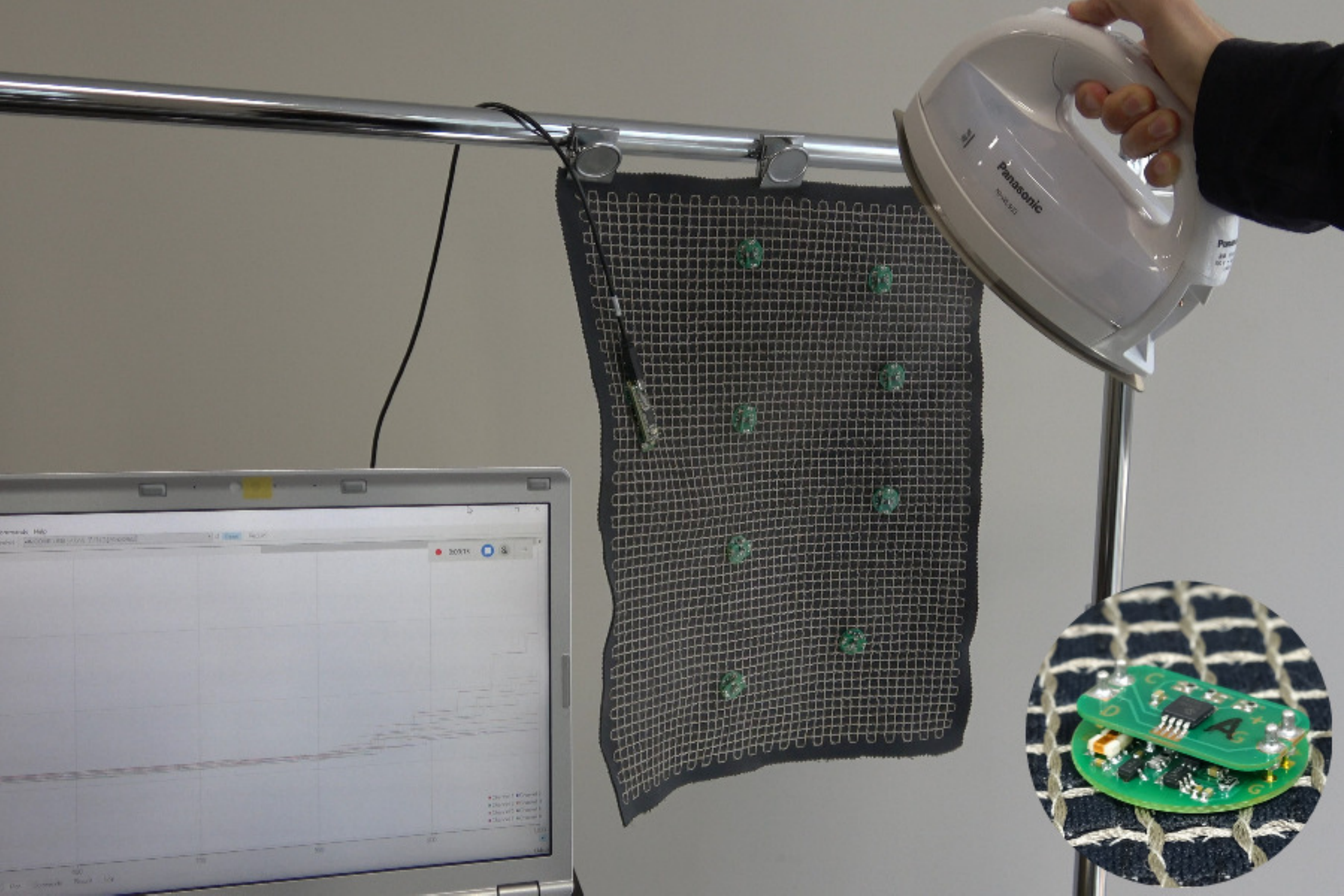}
		\put(0,0){\includegraphics[width=0.98\cw]{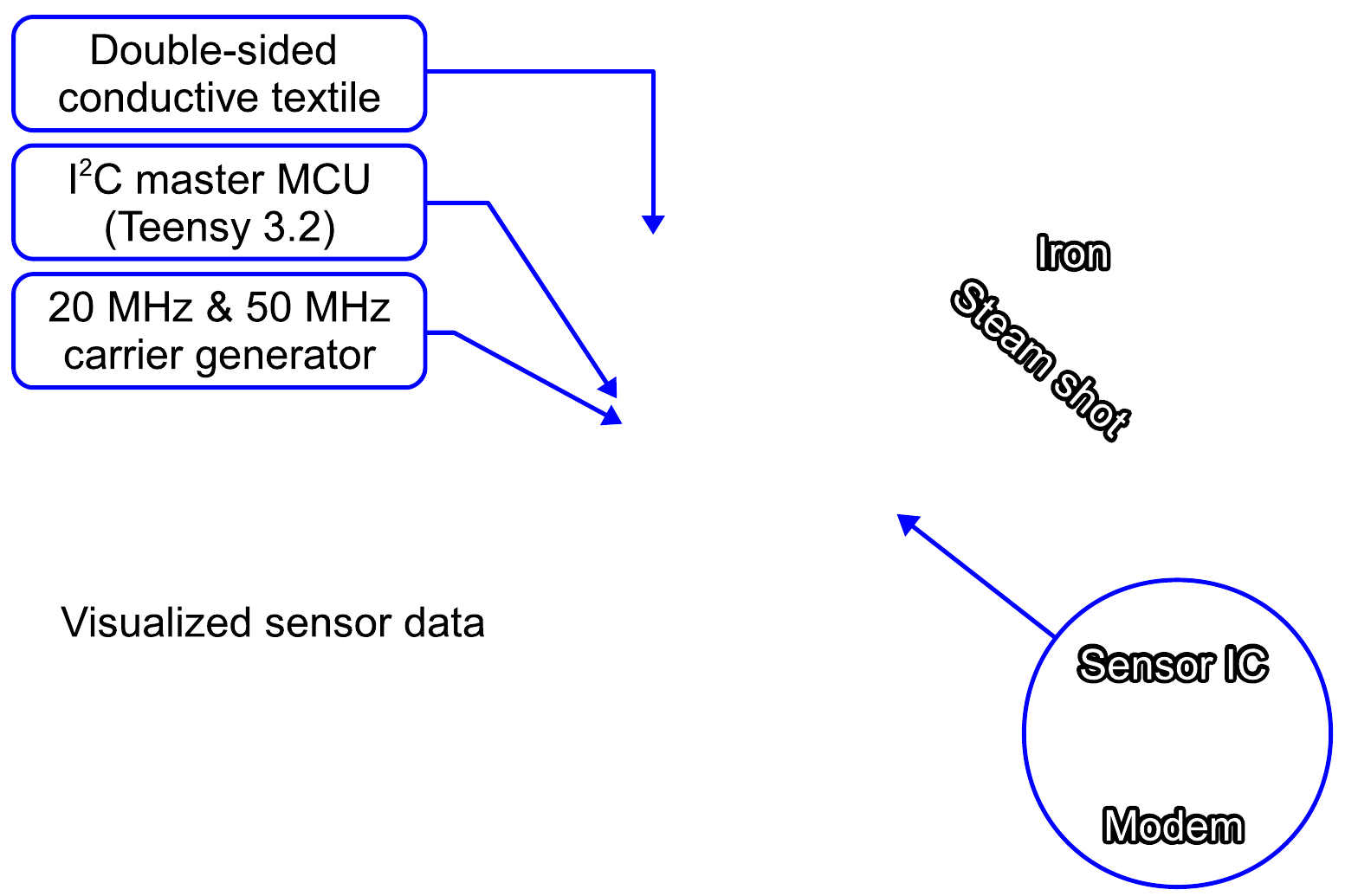}}
	\end{overpic}
	\caption{Demonstration system with distributed temperature sensors {(from A to H)}
	on a double-sided conductive textile with a dimension of 40 cm $\times$ 30 cm.
	{The system captures the temperature change 
	when steam is sprayed onto the textile from an iron.}}
	\label{fig:demo}
\end{figure}

This section presents a demonstration system. 
The master device collects sensor readings from multiple sensor nodes, and the data are visualized on a PC, as shown in \reffig{demo}.
A brief report on the same system has been published at a previous conference \cite{nodaBSN2018}.
\begin{figure}[t]
	\centering
	\resizebox{0.99\cw}{!}{\input{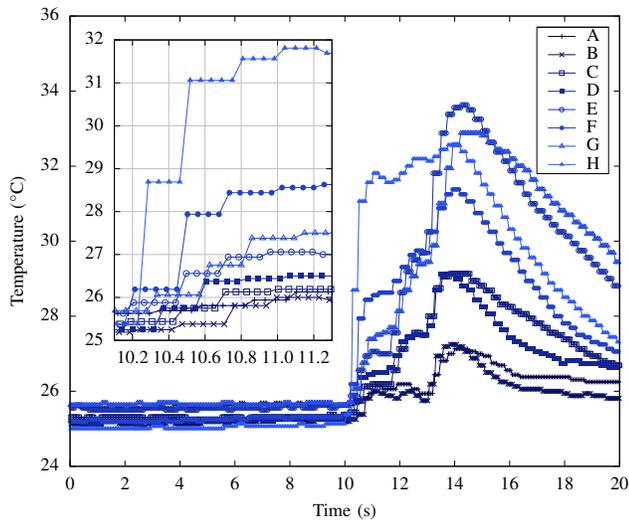}}
	\caption{	{Example of measurement results with a demonstration system
	shown in \reffig{demo}.
	The plot is steppy, especially when the temperature change 
	became steeper, because the sampling frequency of the sensor, 
	4 samples/s, is 
	lower than the \iic\ data transmission frequency, which is
	approximately 17 times/s for each sensor.}}
	\label{fig:8sensor}
\end{figure}

A batteryless sensor node can be composed of the modem module shown in \reffig{modemboard} and a commercially available \iic-enabled sensor IC that operates as an \iic\ slave.
A master device that initiates the \iic\ communication and collects sensor data, can be composed of the modem module and an MCU board. 
We used Teensy 3.2 and a temperature sensor IC, Microchip MCP9808 \cite{mcp9808}, as the master and the slaves, respectively. 
{%
MCP9808 contains eight patterns of user-selectable \iic\ slave addresses. 
Therefore, eight sensors can operate on the same textile at the most.
If the \iic\ slave address is fully programmable, the master can distinguish 112 slaves with the standard 7-bit \iic\ address except for 16 reserved addresses \cite{semiconductorsum10204}. 
}

The entire system is the same as \reffig{i2w}. 
A 3.3-V dc voltage was supplied to the bus via a 47-$\upmu$H RFC inductor. 
Further, 20-MHz and 50-MHz carriers were generated with a clock generator, Silicon Laboratories Si5351A \cite{si5351}. 
The output clock signal is not a pure tone but a rectangular wave containing harmonics. 
To suppress the harmonics, the carries were supplied via LC series resonant circuits.
Each of the pull-up impedances $\Zpo$ and $\Zpt$ is the series connection of the LC filter and a resistor. 
$\Zpo$ is a series connection of a 3.3$\mathchar`-\upmu$H inductor, a 22-pF capacitor, and a 2.0-k$\Omega$ resistor. 
$\Zpt$ is a series connection of a 3.3$\mathchar`-\upmu$H, a 2.2-pF capacitor, and a 2.0-k$\Omega$ resistor. 

The master MCU board was connected to a PC with a universal serial bus (USB) cable, and the collected temperature data were visualized by a plotting software. 
{Time-series temperature data captured when steam is sprayed onto the sensor array is plotted in \reffig{8sensor}.}

Because the \iic\ slave addresses are preset, when the batteryless slaves are attached on the textile, they can immediately join the network, i.e., plug-and-play operation is achieved. 
We have included a supplementary MP4 video clip that demonstrates the plug-and-play operation of the sensor nodes.
This will be available at http://ieeexplore.ieee.org.

\section{Conclusion}
\label{sec:conclusion}

We proposed the \iiw that enables an \iic-formatted signal transfer along with a continuous dc power supply on a single transmission line. 
Double-sided conductive textiles can be used as the transmission line. 
\iiw\ enables the simple implementation of a body sensor network, in which a number of batteryless and antennaless sensors are distributed over clothing. 
Individual wires connecting individual devices are eliminated. 
Each device was attached on the textile medium with a tack connector that allows for the physical mounting and electrical connection to be established using a single action.

The key component for the \iiw\ implementation was the T-network filters that enabled the passive modulation of carriers with SCL and SDA signals. 
The passive modulation approach enabled the carrier generators to be separated from each device, and contributed to the smaller footprint and lower power consumption of each device. 
This paper described the design procedure and operation of the T-filters in detail. 
The reactance of each branch of the T-network was determined by considering the impedance poles and zeros. 

\iic\ is the de facto standard of serial communication between sensor ICs and MCUs, and various sensor ICs equipped with the \iic\ interface are commercially available. 
Additionally, the software libraries of MCUs for \iic\ communication with sensor ICs can be used for \iiw\ without any modification. 
Therefore, the proposed \iiw\ is expected to be used widely for wearable sensor systems. 

\ifCLASSOPTIONcaptionsoff
\newpage
\fi


\end{document}